\pdfoutput=1 
\RequirePackage{fix-cm}
\documentclass{svjour3}                     
\smartqed  
\usepackage[caption=false]{subfig}
\usepackage{floatrow}
\usepackage{graphicx}
\usepackage{natbib,amsfonts,amssymb}
\usepackage[section]{placeins}
\usepackage{times} 
\begin{document}

\title{Retrograde resonance in the planar three-body problem 
}


\author{M. H. M. Morais       \and
         F. Namouni 
}


\institute{M. H. M. Morais \at
              Department of Physics \& I3N, University of Aveiro, Campus de Santiago, 3810-193 Aveiro, Portugal \\
                 \email{helena.morais@ua.pt}           
           \and
           F. Namouni \at
              Universit\'e de Nice, CNRS, Observatoire de la C\^ote d'Azur, BP 4229, 06304 Nice, France
}

\date{Received: date / Accepted: date}

\maketitle

\begin{abstract}
We continue the investigation of the dynamics of retrograde resonances initiated in \citet{Morais&Giuppone2012}. After deriving a procedure to deduce  the retrograde resonance terms from the standard expansion of the three-dimensional disturbing function, we concentrate on the  planar problem and construct surfaces of section that explore phase-space in the vicinity of the main retrograde resonances (2/-1, 1/-1 and 1/-2). In the case of the 1/-1 resonance for which the standard expansion is not adequate to describe the dynamics, we develop a semi-analytic model  based on numerical averaging of the unexpanded disturbing function, and show that the predicted libration modes are in agreement with the behavior seen in the surfaces of section.
\keywords{Resonance. Three-body problem. Surface of section.}
\end{abstract}

\section{Introduction}
The discovery of extrasolar planets that orbit their host stars in the direction opposite the star's rotation has renewed interest in the dynamics of retrograde motion in gravitational systems \citep{Triaud2010}. In the solar system, retrograde motion is confined to smaller bodies such as satellites of the outer planets and long period comets. Understanding the structure of retrograde motion and in particular retrograde resonances will help elucidate the origin and  evolution of the observed systems. \citet{Gayon&Bois2008} performed numerical integrations of systems with planets moving in opposite directions and observed that  a retrograde resonance in two-planet systems is more stable than the equivalent prograde resonance confirming the idea that when two bodies orbit in different directions, encounters  occur at a higher relative velocity during a shorter time and mutual perturbations are therefore weaker. In \citet{Gayon_etal2009CMDA} the authors obtained the 2/1 retrograde resonance disturbing function, and identified retrograde resonance angles. However, they were unable to find initial conditions that correspond to libration in retrograde resonance, and did not identify the theoretical reason for the observed enhanced stability of retrograde resonances. In \citet{Morais&Giuppone2012} (hereafter Paper I) we compared the stability of prograde and retrograde planets within a binary system. We observed that retrograde planets remain stable nearer to the secondary star than  prograde planets. We showed that instability is caused by  single mean motion resonances (MMRs) and the possible overlap of adjacent pairs.  We used a standard expansion of the disturbing function for the planar circular restricted three-body problem (CR3BP) to obtain the retrograde resonance terms, and we explained  how these terms show why retrograde resonances are more stable than prograde resonances. Indeed, the magnitude of the p/q resonance terms is proportional to a power of the  eccentricity, which at the lowest order, is $e^{p+q}$ in the retrograde case and $e^{p-q}$ in the prograde case.

In this paper, we continue our investigation of retrograde motion in the three-body problem by systematically studying the structure of the phase space near the main retrograde resonances. We concentrate on the planar problem and examine in detail motion near the 2/1, 1/1 and 1/2 retrograde resonances. In section 2,  we explain how to obtain the retrograde resonance terms from an expansion in Laplace coefficients of the  three-dimensional disturbing function. We also  show that, in the planar problem, these coincide with the retrograde resonance terms obtained in Paper I.  Since Laplace coefficients diverge when  the semi major axes' ratio is close to unity,  in section 3, we develop a semi-analytical model for the co-orbital 1/1 retrograde resonance based on numerical averaging of the unexpanded disturbing function. In section 4, we present our numerical approach and results, and describe how retrograde motion phase-space  is structured and where stable motion is possible. Section 5 contains a discussion of our results. 

\section{Differences between prograde and retrograde resonance}

The encounter of two bodies in a retrograde configuration (orbiting in opposite directions) occurs at a higher relative velocity during a shorter time than in a prograde configuration. This implies that mutual perturbations are  generally weaker for retrograde MMRs. In Paper I, we studied retrograde MMRs analytically  in the context of the planar circular restricted three-body problem and compared their relative strength and stability to prograde  resonances. Here, we will explain how we obtain the slow terms of the disturbing function for a p/q retrograde MMR in the three-dimensional CR3BP.

\subsection{Disturbing function}

Consider a test particle that moves under the gravitational effect of a binary composed of a primary with mass $M_{\star}$ and a secondary with mass $m\ll M_{\star}$. The motion of $m$ with respect to $M_{\star}$ is a circular orbit of  radius $a^\prime=1$ and longitude angle $\lambda^\prime$.  The reference plane is defined by the binary's orbit. The test particle's osculating Keplerian orbit with respect to $M_{\star}$  has semi-major axis $a$, eccentricity $e$, inclination $I$, true anomaly $f$, argument of pericentre $\omega$, and  longitude of ascending node $\Omega$.  The  disturbing function  reads:
\begin{equation}
R= G\,m (1/\Delta-r\,\cos{\psi})
\end{equation}
where $r$ is the radius of test particle, 
 $\psi$ is the angle between the radius vectors of the binary and the test particle,
$\Delta^2=1+r^2-2\,r\,\cos{\psi}$, and
\begin{equation}
\cos\psi=\cos(\Omega-\lambda^\prime)\cos(\omega+f)-\sin(\Omega-\lambda^\prime)\sin(\omega+f)\cos I. 
\end{equation}
The first term of $R$ (direct perturbation) is the gravitational force from the mass $m$ on the test particle whereas the second term (indirect perturbation) comes from the reflex motion of the star under the influence of the mass $m$ as the standard coordinate system is chosen to be  centered on the star.

The classic series of the disturbing function is expanded in powers of $\sin^2(I/2)$. This is adequate for nearly coplanar prograde motion since $\sin^2(I/2)\approx 0$ but not for nearly coplanar retrograde motion since $\sin^2(I/2)\approx 1$. Therefore, for nearly coplanar orbits, we define $\beta\ll 1$ such that $I=\beta$ or $I=180^\circ-\beta$, respectively, for prograde or retrograde motion. We may therefore write:
\begin{equation}
\cos\psi = (1-s^2) \cos(f+\omega\pm(\Omega-\lambda^\prime)) +  s^2 \cos(f+\omega \mp(\Omega-\lambda^\prime)),
\end{equation}
where $s^2=\sin^2(\beta/2)\ll 1$, and the $\pm$ sign applies to prograde or retrograde cases, respectively.

Next, we write $\Delta^2=1+r^2-2\,r\,\cos(f+\omega\pm(\Omega-\lambda^\prime))-2r\Psi$, where  $\Psi$ is defined as:
\begin{equation}
\Psi = \cos{\psi}-\cos(f+\omega\pm(\Omega-\lambda^\prime))     = 2\, s^2 \sin(\pm(\Omega-\lambda^\prime))\sin(\omega+f) .\label{PSI}
\end{equation}
Expanding the direct perturbation term $\Delta^{-1}$ in the vicinity of $\Psi=0$ (as $s^2\ll 1$),  we may write:
\begin{eqnarray}
\label{dfunction}
\frac{1}{\Delta} &=& \sum_{i=0}^{\infty} \frac{(2 i) !}{(i !)^2} \left( \frac{1}{2} r \Psi \right)^{i}  \frac{1}{\Delta_{0}^{2\,i+1}},
\end{eqnarray}
where $\Delta_0^2=1+r^2-2\,r\,\cos(f+\omega\pm(\Omega-\lambda^\prime))$.

Finally, defining  $\epsilon=r/a-1=\mathcal{O}(e)$ and expanding $\Delta_{0}^{-(2\,i+1)}$ around $\epsilon=0$:
\begin{eqnarray}
\frac{1}{\Delta_{0}^{2\,i+1}} &=& 
\left( 1+\sum_{k=1}^{\infty}\frac{1}{k!}\,\epsilon^k\,\alpha^k\,\frac{d^k}{d_k \alpha} \right) \frac{1}{\rho^{2\,i+1}},
\end{eqnarray}
 with $\alpha=a/a^{\prime}$, and
\begin{eqnarray}
\frac{1}{\rho^{2\,i+1}}&=& (1+\alpha^2-2\,\alpha\,\cos(f+\omega\pm(\Omega-\lambda^\prime)))^{-(i+1/2)}, \nonumber \\
 &=& \sum_{j} \frac{1}{2} b_{i+1/2}^{j}(\alpha) \cos(j (f+\omega\pm(\Omega-\lambda^\prime))), \label{RHO}
\end{eqnarray} 
where $b_{i+1/2}^{j}(\alpha)$ are Laplace coefficients. For   $\alpha<1$, they may be expanded as convergent series in $\alpha$ \citep{Ellis&Murray2000}.

Therefore, for retrograde orbits ($I\approx 180^\circ$) the disturbing function (Eq.~\ref{dfunction}) is expanded in powers of $\cos(I/2)\ll 1$, whereas for prograde orbits ($I\approx 0$) it is expanded in powers of $\sin(I/2)\ll 1$. \citet{Yokoyama_etal2005} studied the effect of Triton's retrograde orbit on the motion of Neptune's satellites. They used computer algebra to expand the disturbing function in powers of $\sin(I)\ll 1$, and compared direct  numerical integrations of the equations of motion with results obtained from the integration of Lagrange's equations using the retrograde disturbing function. The advantage of our approach is that it allows us to obtain the retrograde disturbing function directly from the well known prograde disturbing function, without the need for specific computer algebra. We will now explain that procedure in detail.

Examination of the angles in the expression of $\Psi$ (\ref{PSI}) and that of $\rho$ (\ref{RHO}) show that there are two ways with which 
 the expansion of the disturbing function for retrograde motion can be obtained from the expansion of the disturbing function for prograde motion. The two ways  are equivalent as they depend on whether one chooses  to invert the motion of the inner body or that of the outer one\footnote{A retrograde orbit  with inclination $I>90^\circ$ can be obtained from a prograde orbit with inclination $180^\circ-I$ by inverting the direction of motion which implies a swap between ascending  and descending nodes.}. Inverting the motion of the outer body  gives the first transformation: $I^\star=180^\circ-I, \ \lambda^{\prime\star}=-\lambda^\prime$, $\omega^\star=\omega-\pi$ and  $\Omega^\star =-\Omega-\pi$. In this case, the longitude of pericentre $\varpi=\omega+\Omega$ is transformed into $\varpi^\star=\omega-\Omega$ and the mean longitude $\lambda=M+\omega+\Omega$ into  $\lambda^\star=M+\omega-\Omega$ where $M$ is the mean anomaly.  Equivalently, this means applying the generating function $F_1=-\lambda^\prime\Lambda^{\prime\star}+(\lambda+2z)\Lambda^\star +(g-2z) \Gamma^\star-(z-\pi) Z^\star$ to the usual Poincar\'e action-angle variables. Inverting the motion of the inner body gives the second possible transformation:  $I^\star=180^\circ-I, \ f^{\star}=-f,\ \omega^\star=\pi-\omega$ and $\Omega^\star =\pi+\Omega$ that may be obtained with the generating function $F_2=\lambda^\prime\Lambda^{\prime\star}-(\lambda+2z)\Lambda^\star -(g-2z) \Gamma^\star +(z-\pi)Z^\star$. We note however that these two transformations are passive in that they allow us only to obtain the expression of the resonant arguments. Once the arguments are obtained formally, the assumption that $\dot\lambda>0$ and $\dot\lambda^\prime>0$ always holds. This is in contrast to the approach adopted in Paper I where an active transformation was used to study the planar dynamical problem by choosing explicitly from the outset the convention $\dot{\lambda}>0$ and $\dot{\lambda^\prime}<0$. 
A similar transformation has been used by \citet{Saha&Tremaine1993} to analyze long-term numerical integrations of the retrograde jovian satellites.

\subsection{Resonant terms}

Now that we have shown how the expansion of the disturbing function  for $I=180^\circ-\beta$ is obtained from that with  $I=\beta$, we may use the literal expansion of \citet{Ellis&Murray2000} valid for prograde motion and transform the relevant resonance terms to describe the corresponding retrograde resonance. We will use the first transformation described above with $s_{\star}=\cos(I/2)$, $\lambda^\star=M+\omega-\Omega$ and $\varpi^\star=\omega-\Omega$.

The 2/1 retrograde resonance terms are of type $e^3\cos(\lambda^\star-2\,\lambda^\prime-3\,\varpi^\star)$ [term 4D3.1 with $j=2$], and  $e\,s_{\star}^{2}\cos(\lambda^\star-2\,\lambda^\prime-\varpi^\star+2\,\Omega)$ [term 4D3.5 with $j=2$].
The 1/2 retrograde resonance has direct and indirect terms of type  $e^3\cos(2\,\lambda^\star-\lambda^\prime-3\,\varpi^\star)$ [terms 4D3.4 with $j=2$ and 4I3.6], and $e\,s_{\star}^{2}\cos(2\,\lambda^\star-\lambda^\prime-\varpi^\star+2\,\Omega)$ [terms 4D3.10 with $j=2$ and 4I3.13]. 
The 1/1 retrograde resonance has direct and indirect terms. These are of type  $e^2\cos(\lambda^\star-\lambda^\prime-2\,\varpi^\star)$ [terms 4D2.1 with $j=1$ and 4E2.2 ], and $s_{\star}^{2}\cos(\lambda^\star-\lambda^\prime+2\,\Omega)$ [terms 4D2.4 with $j=1$ and 4E2.6]. 
However, the  1/1 resonance direct terms  cannot be obtained from the literal expansion of the disturbing function (since Laplace coefficients diverge when $\alpha\to 1$). We develop a semi-analytic model for the co-orbital resonance in the next section.

A similar analysis for any $p/q$ retrograde resonance shows that  there are resonant terms\footnote{D'Alembert rule is not obeyed because the canonical transformations described in Sect.~2.1 imply that angles  for the test particle are measured  in the opposite direction of the binary's motion.}   
\begin{equation}
e^{p+q-2\,k}\,s_{\star}^{2\,k}\cos(q \lambda^\star- p \lambda^\prime-(p+q-2\,k)\varpi^\star +2\,k \Omega) . 
\end{equation}
with $k=0,1,2,...$ and $p+q\ge2\,k$.

If $\beta\ll 1$ then $s_{\star}\ll1$, hence the term with $k=0$ is dominant. Since we restrict our study to planar retrograde resonance ($s_{\star}=0$),  only the term $e^{p+q}\cos\phi$ remains, where
\begin{equation}
\phi= q \lambda^\star- p \lambda^\prime-(p+q)\varpi^{\star} . 
\end{equation}
We thus recover the retrograde resonant angle from Paper I (with the expected  change of sign for the term in $\lambda^\prime$). Following Paper I, we use the notation p/-q resonance when referring to a p/q retrograde resonance.

\section{A model for co-orbital resonance}
Consider a test particle in the co-orbital region of the secondary ($|a-1|\ll 1$). The disturbing function may be expressed using the natural angles: 
the fast epicyclic motion represented by the mean longitude $\lambda$ of the
particle   and the guiding centre phase represented by the relative mean longitude $\tau =\lambda-\lambda^\prime$.  To obtain the resonant Hamiltonian, the disturbing function, $R$ (Eq.~1), is averaged with respect to the fast angle $\lambda$. The corresponding function is the  ponderomotive potential $S=\langle R \rangle$ used in our previous work on the co-orbital resonance \citep{Namouni1999,Namouni_etal1999}. When the relative longitude $\tau$ is introduced, we may write:
\begin{equation}
\cos\psi= \frac{1}{2}\, (1+\cos I) \cos(f-M+\tau)+\frac{1}{2}\, (1-\cos I) \cos(f+M-\tau+2\omega),
\end{equation}
and the ponderomotive potential is given as:
\begin{equation}
S= \frac{1}{2\pi(1-e^2)^{1/2}}\int_0^{2\pi} R  \,  r^2 df,
\end{equation}
where $r=a (1-e^2)/(1+e\cos f)$, and  the average over the mean anomaly $M$ has been replaced by an average over $f$ using the conservation of angular momentum. The mean anomaly $M$ is related to the true anomaly by the eccentric anomaly $\tan (E/2) =(1-e)^{1/2}(1+e)^{-1/2}\tan (f/2)$ and Kepler's equation $M=E-e \sin E$. 

\begin{figure*}[b]
\includegraphics*[width=4.1cm]{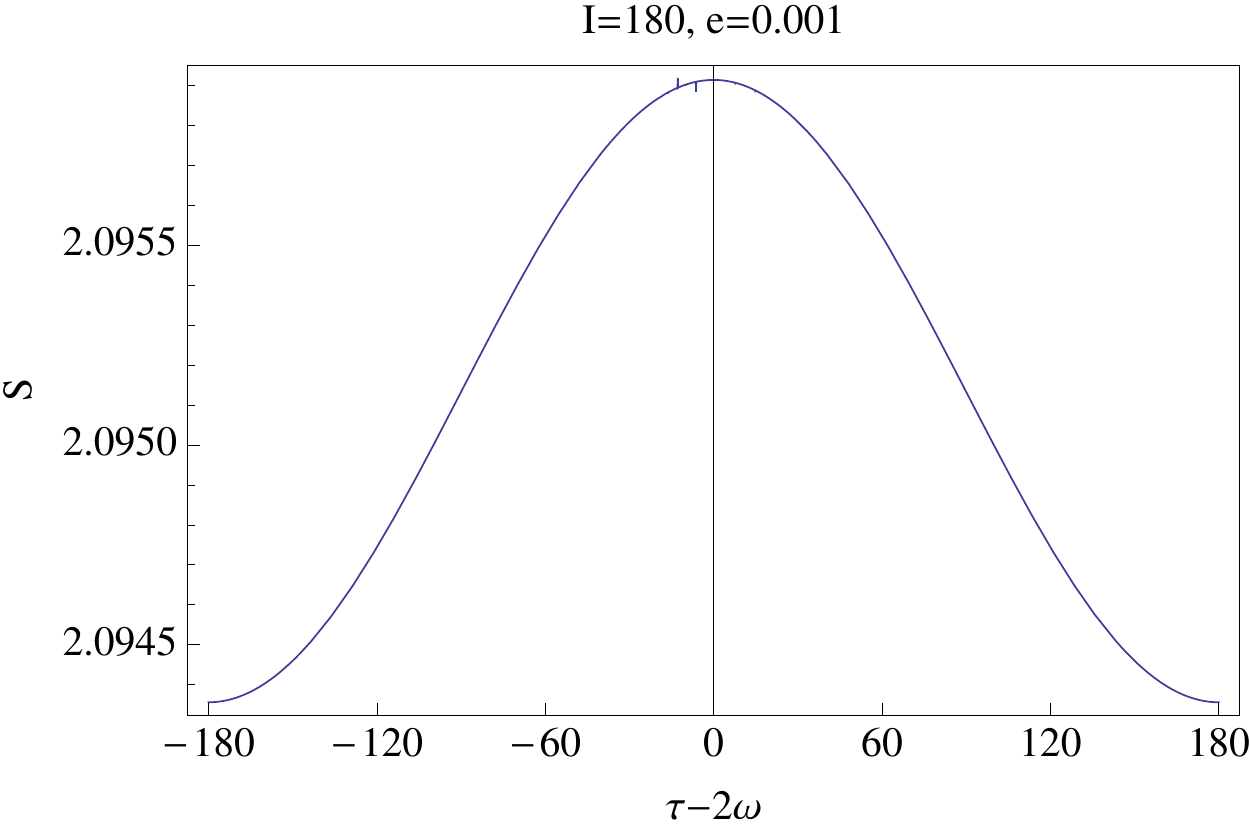}\includegraphics*[width=4.1cm]{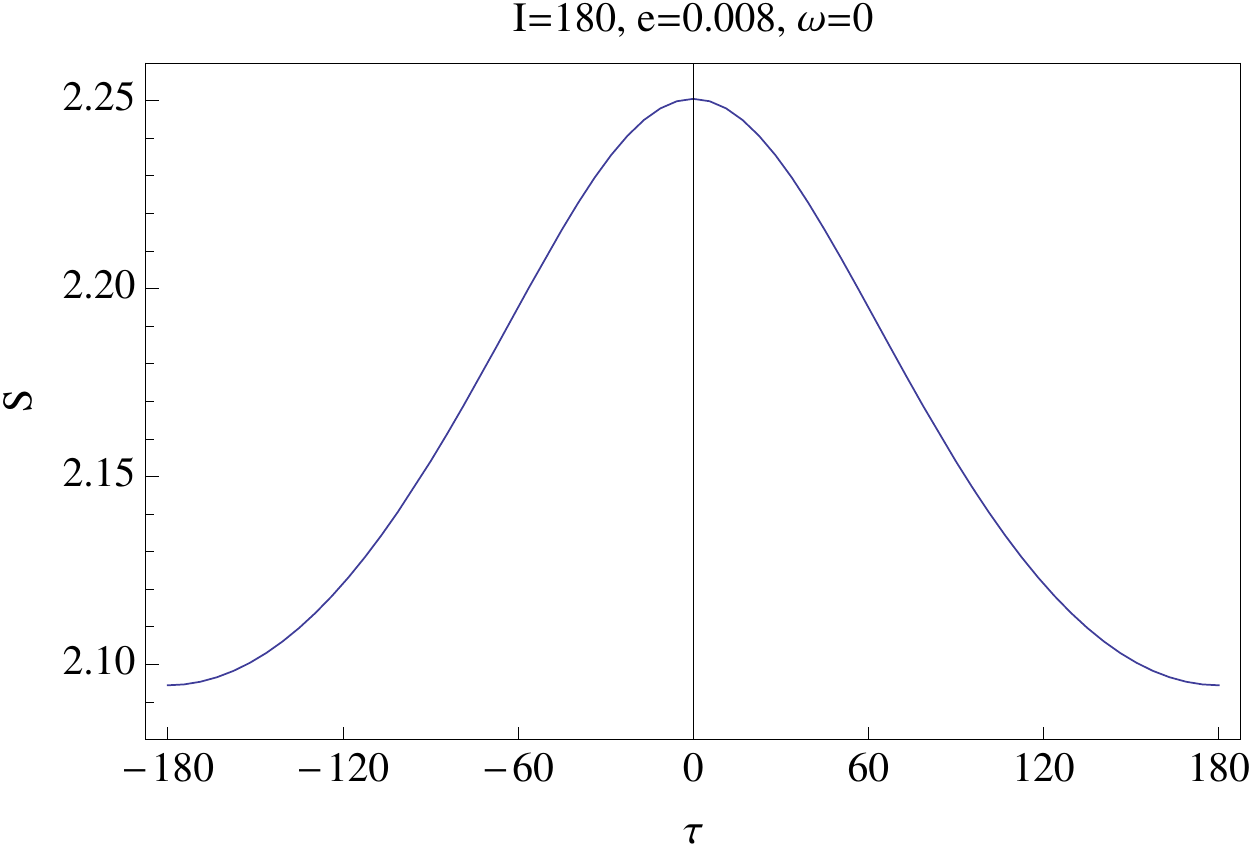}\includegraphics*[width=4.1cm]{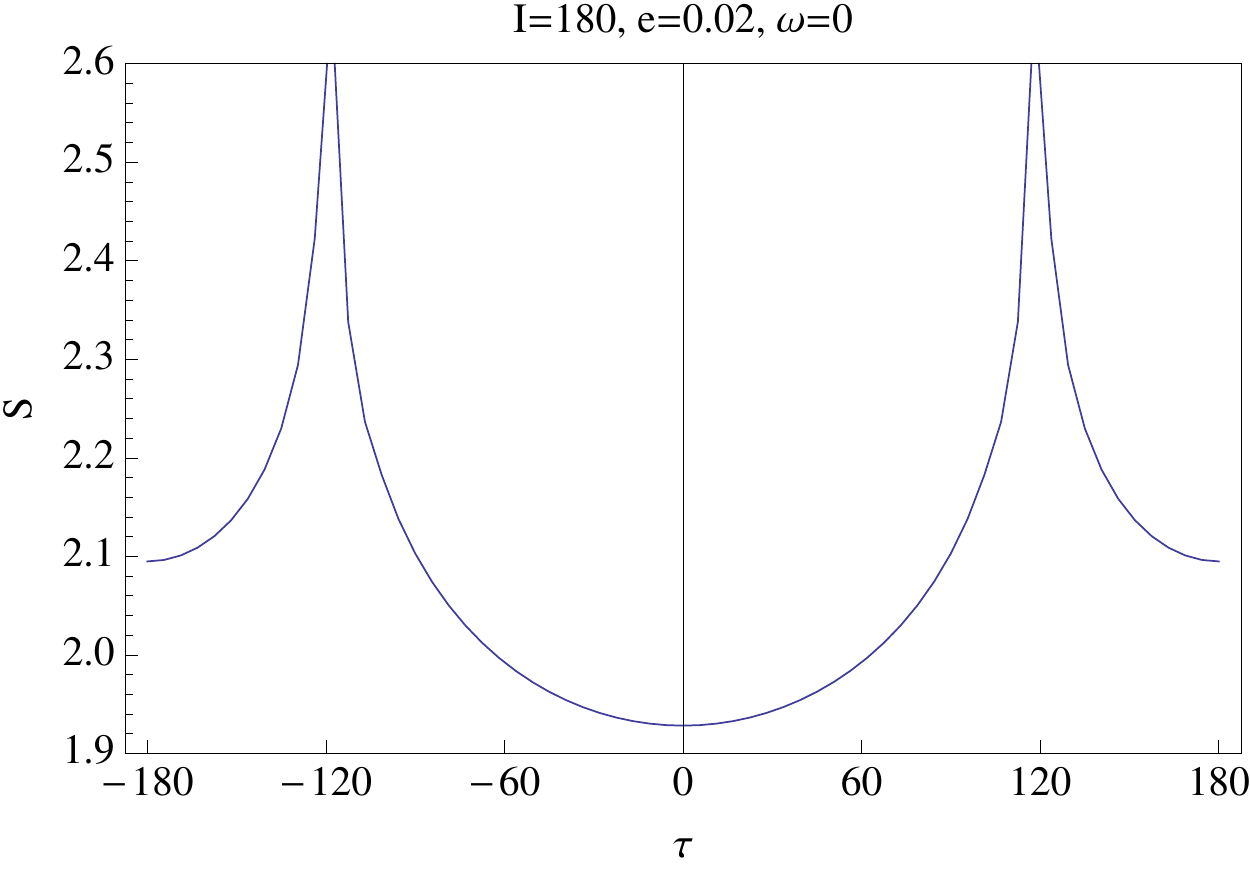} \\ 
\includegraphics*[width=4.1cm]{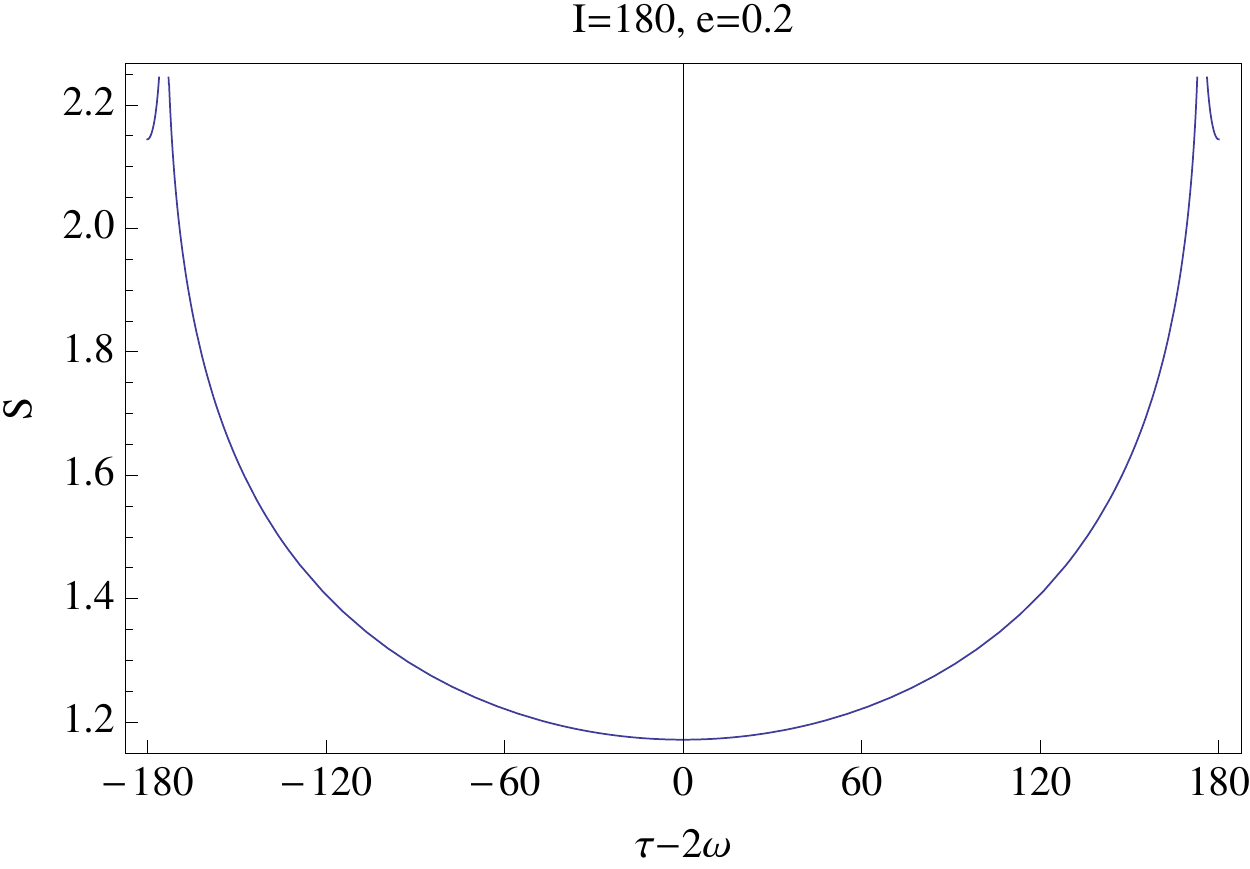}\includegraphics*[width=4.1cm]{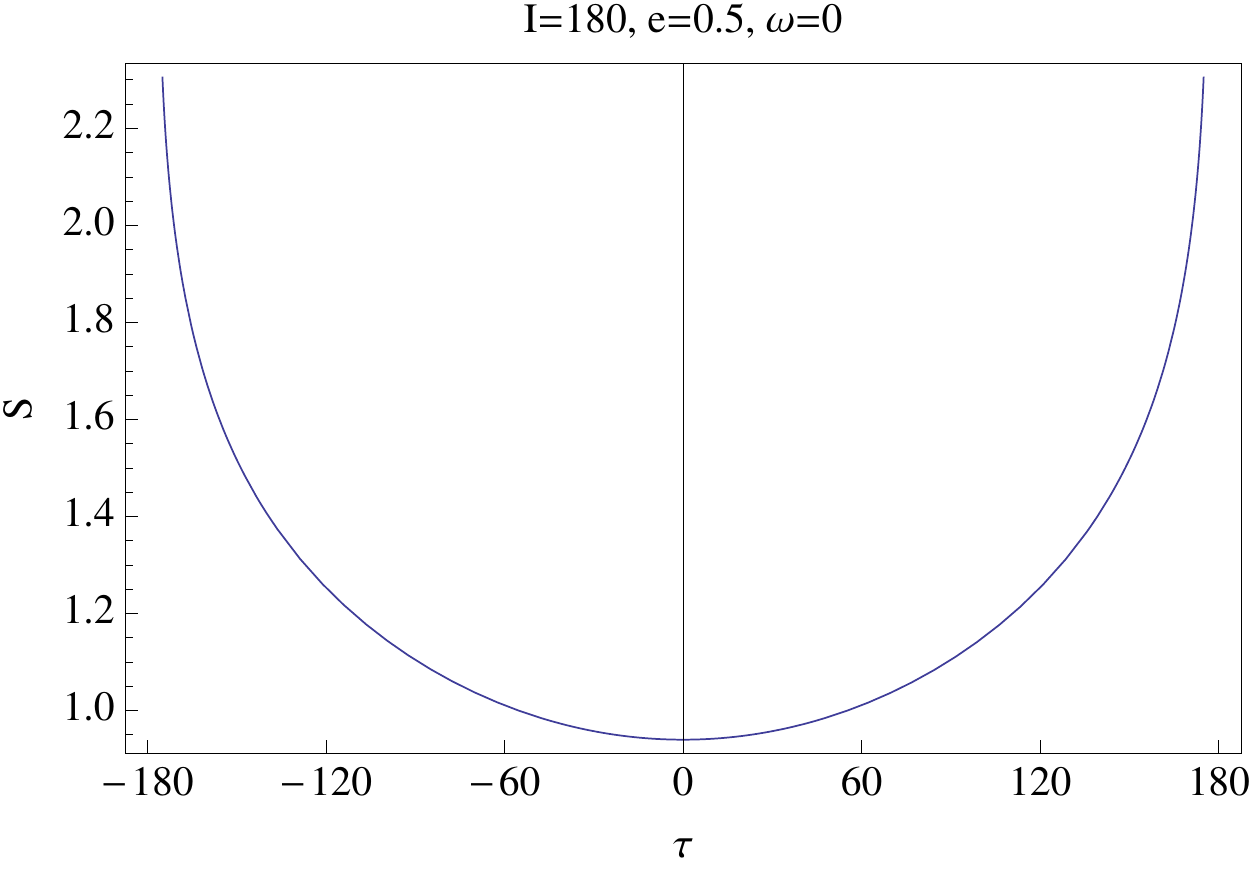}\includegraphics*[width=4.1cm]{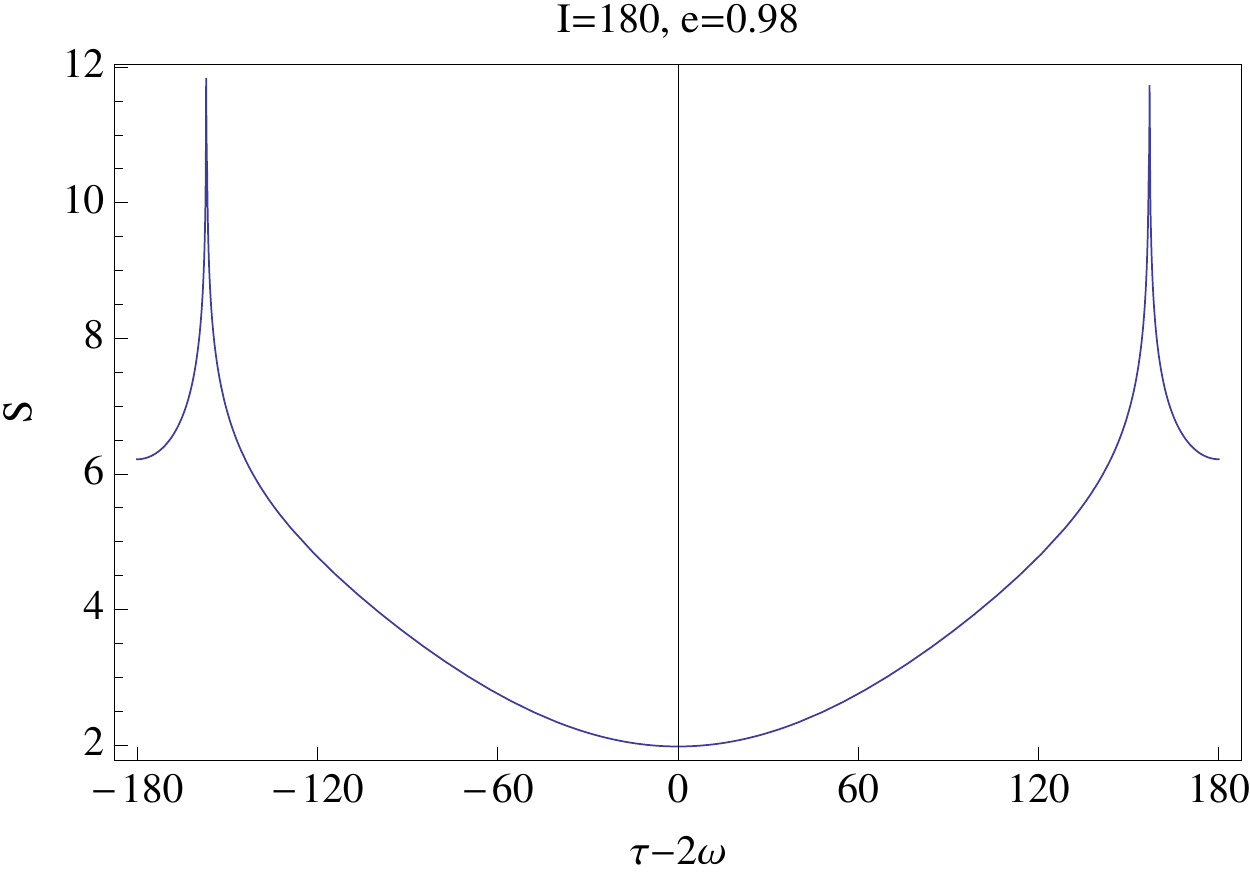} \\ 
\includegraphics*[width=4.1cm]{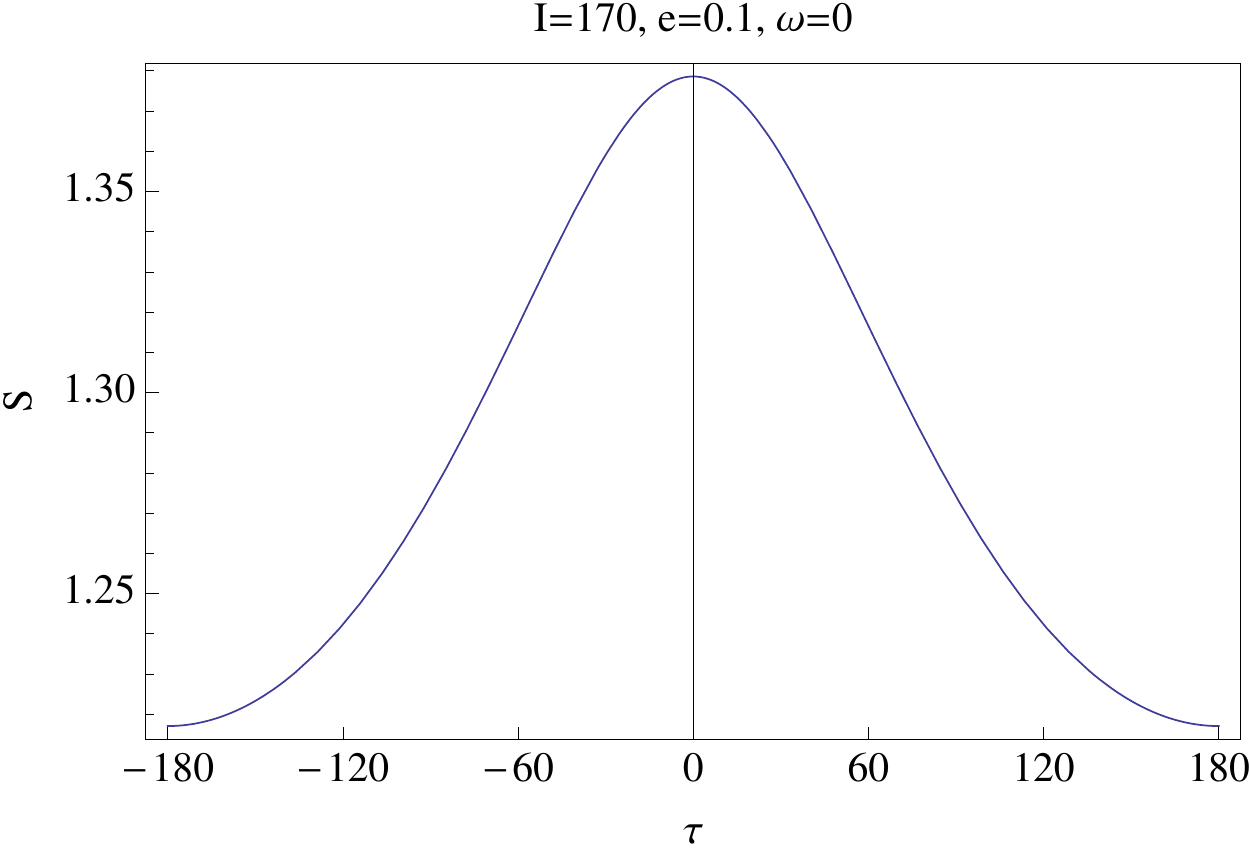}\includegraphics*[width=4.1cm]{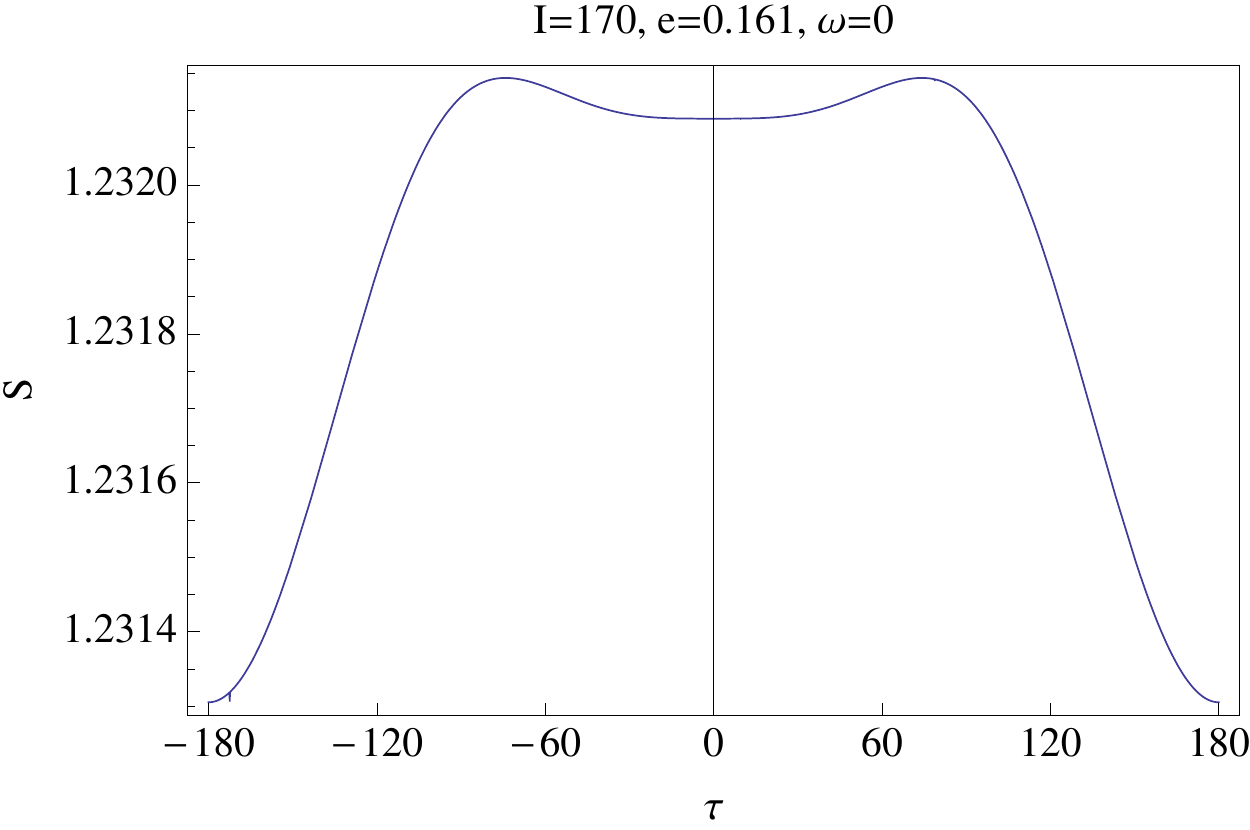}\includegraphics*[width=4.1cm]{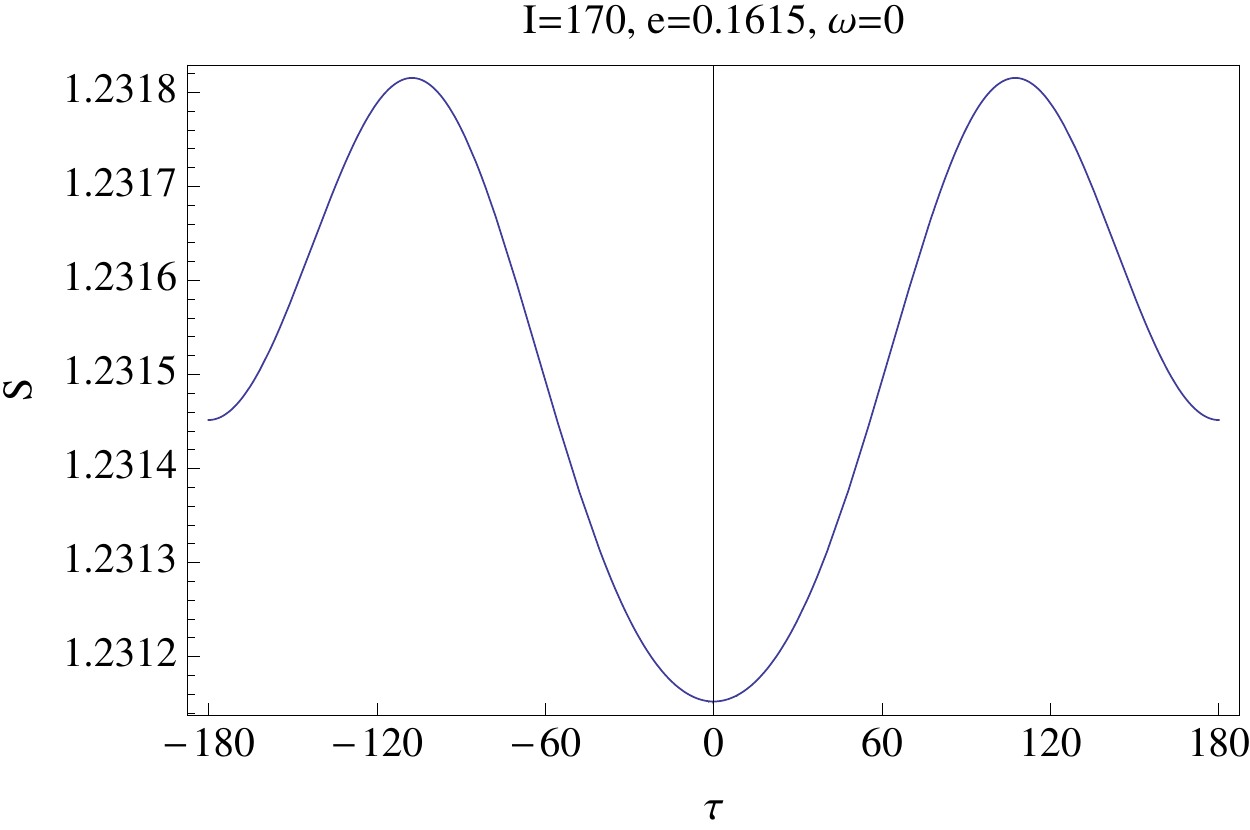} \\ 
\includegraphics*[width=4.1cm]{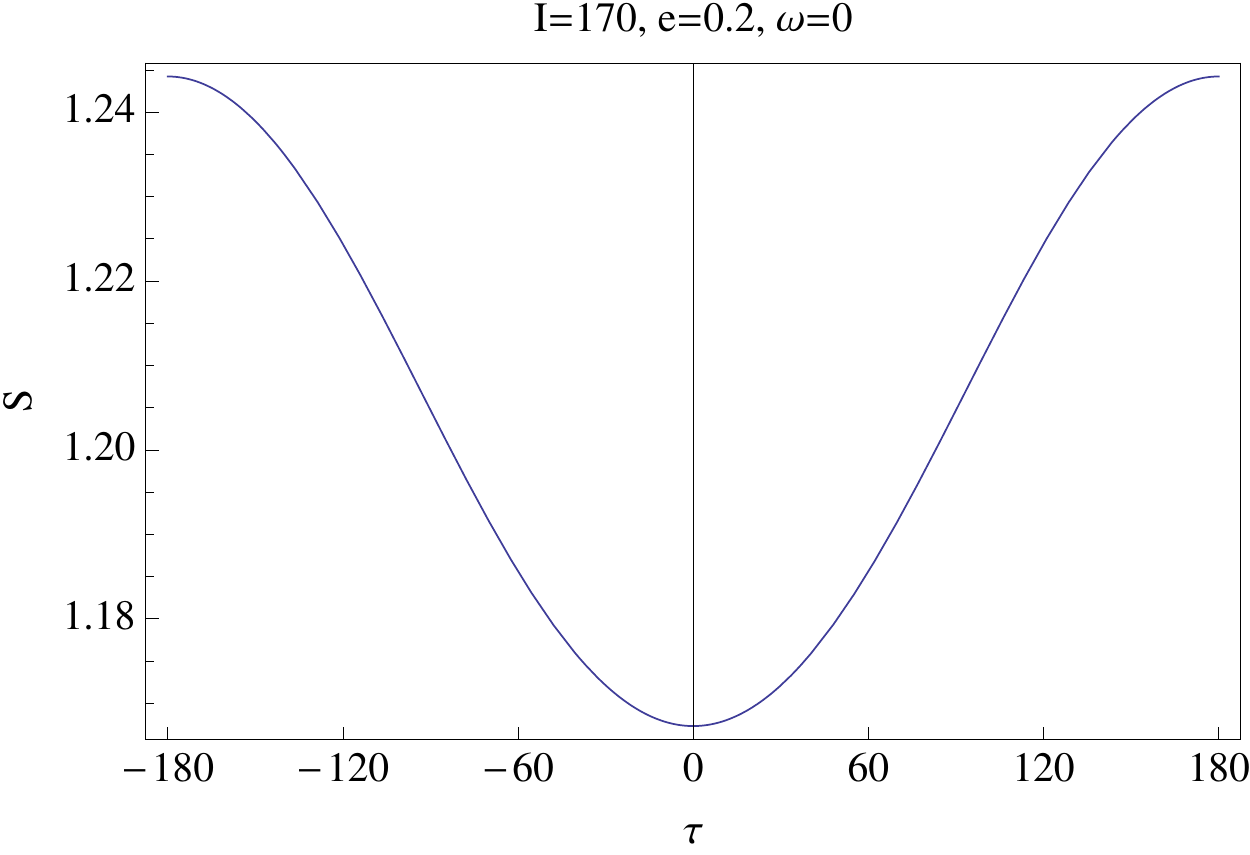}\includegraphics*[width=4.1cm]{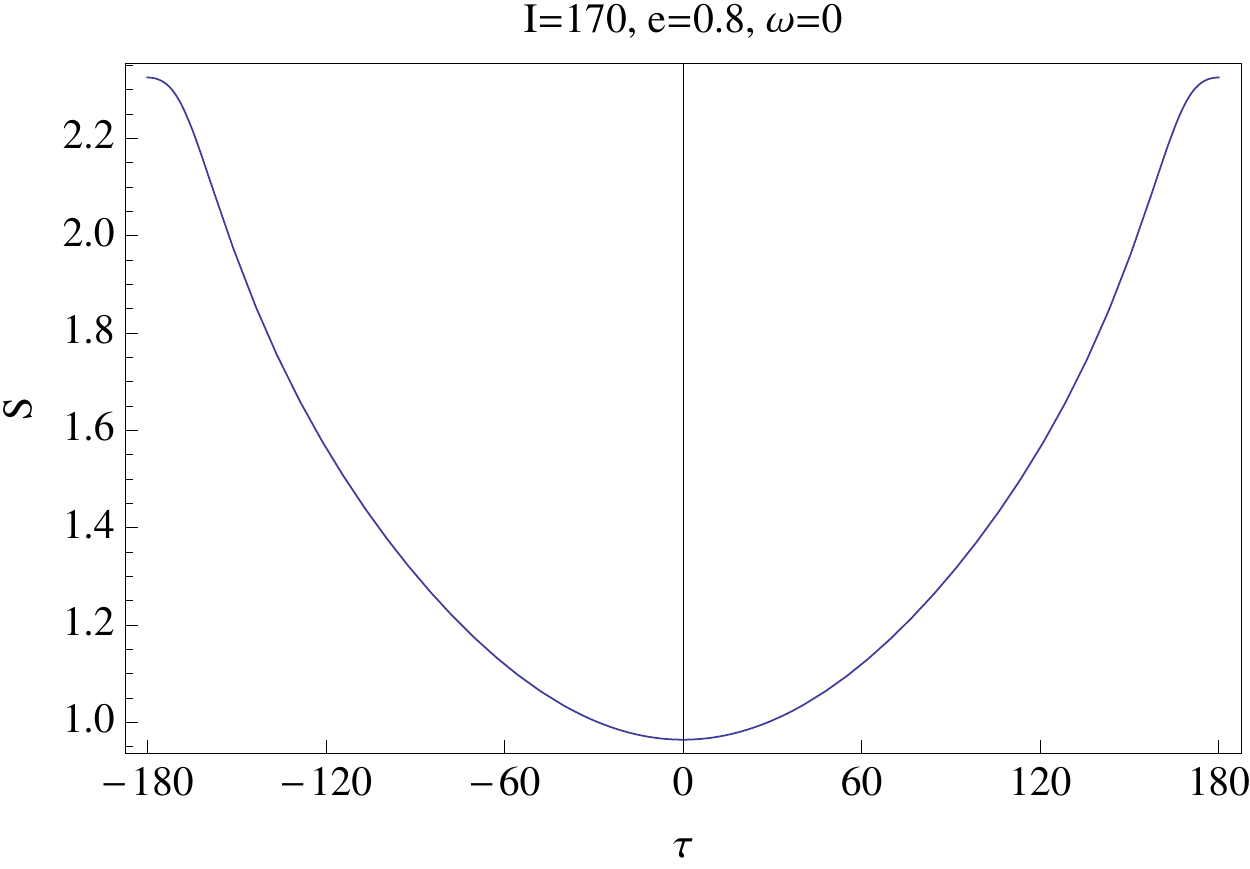}\includegraphics*[width=4.1cm]{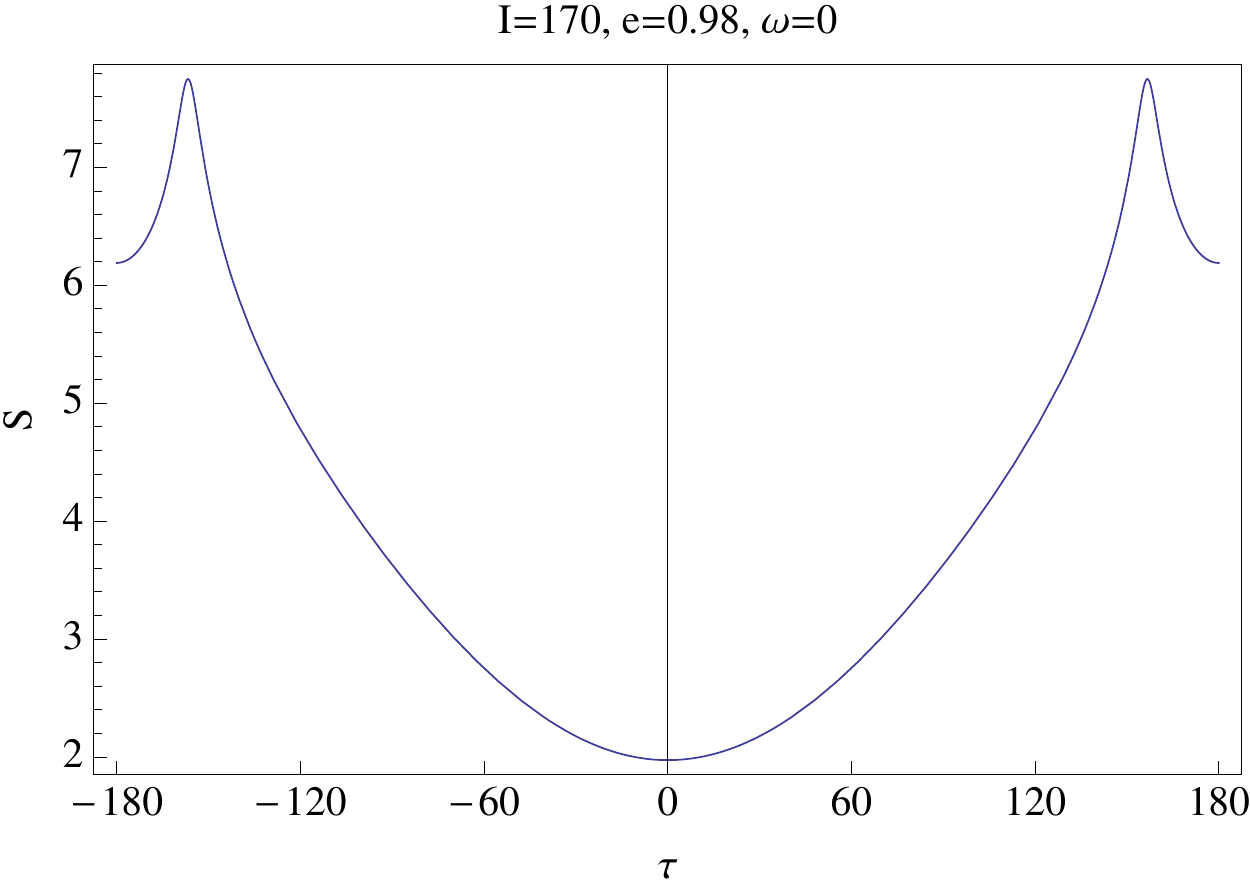} 
\caption{Ponderomotive potential $S$ of the 1/-1 resonance as a function of the resonant angle. The relative semi-major axis is $0.01$. The 1st and 2nd rows illustrate the planar potential whereas the 3rd and 4th rows apply to three dimensional orbits with small inclination. For the latter, $\tau-2\omega$ is no longer the only possible resonant argument and $S$ is plotted as a function of $\tau$ for a fixed $\omega$.}    
\label{figfat1}
\end{figure*}

The expansion-free expression of $\psi$ gives the natural resonant angles for planar motion. For prograde motion ($\cos I=1$), libration occurs around $\phi=\tau$ whereas for retrograde motion ($\cos I=-1$),  libration occurs around $\phi=\tau-2\omega$\footnote{Here, we define $\lambda=M+\omega+\Omega$. If we define $\lambda^\star=M+\varpi^\star$ with $\varpi^\star=\omega-\Omega$ then the retrograde resonant angle is $\phi=\lambda^\star-\lambda^\prime-2\,\varpi^\star$ in agreement with the conclusions of the previous section.}. Figure \ref{figfat1}  (1st and 2nd rows) shows the shape of the potential $S$ as a function of the resonant argument $\phi=\tau-2 \omega$ for planar retrograde motion where $a -1=0.01$.  At low eccentricity, libration occurs only around $\phi=180^\circ$ and the potential is quite shallow. This explains why in the next section we observe that low eccentricity libration orbits for relatively large mass ratios (e.g. $\mu =0.01$) are difficult to set up as the larger the mass ratio the stronger the mutual perturbations, the more destructive the close encounters. We shall show that such librations are quite stable at smaller mass ratios. As the eccentricity is increased, the collision boundary appears and librations may occur around 0 or $180^\circ$. The extent of the libration amplitude depends on eccentricity, for $0.1\lesssim e\lesssim 0.7$, libration around zero has the largest amplitude. Libration around $180^\circ$ regains some importance as $e$ approaches unity. As the ponderomotive potential $S$ is derived for three-dimensional orbits, it is instructive to see how the introduction of a small inclination modifies the dynamics as realistic orbits in the planetary three-body problem never lie exactly on the same plane. Moreover, in the co-orbital resonance, inclination is known to mitigate collisional encounters  and facilitate stable orbital transitions  \citep{Namouni1999,Namouni_etal1999}.  Figure \ref{figfat1} (3rd and 4th rows) shows how $S$ is modified when the retrograde orbits have a mutual inclination of $10^\circ$ and $\omega$ is set to zero. As expected, collision singularities are absent. There appears a bifurcation near $e=0.161$ where librations around 0 and $180^\circ$ have comparable amplitudes and may be associated with the same energy level. The remaining features of the planar problem are present: the potential's shallowness for low eccentricity and the dominance of libration around zero for more eccentric orbits.  We remark that the potential's amplitude and bifurcation modes depend on the relative semi-major axis. Figure \ref{figfat1} only illustrates the similarities and differences with the planar problem. We also note that in the fully three dimensional problem, the time evolution of the argument of pericenter modifies the potential's shape and equilibria whereas for planar orbits, the potential depends on the combined phase $\phi=\tau-2\omega$. We shall  present the study of the three-dimensional retrograde co-orbital resonance elsewhere.

\section{Retrograde resonance phase space}
Poincar\'e surfaces of section are useful tools to study the phase space structure in the three-body problem. In what follows, we define how we set up surfaces of section for the phase space of the 2/-1, 1/-1 and 1/-2 resonances. We then examine the types of orbits involved, as well as their potential stability.

\subsection{Surface of section construction}
In the barycentric rotating frame, the planar circular restricted three-body problem has two degrees of freedom $(x,y)$ and one integral of motion, the Jacobi constant \citep{ssdbook}:
\begin{equation}
C=x^2+y^2-(\dot{x}^2+\dot{y}^2)+\frac{2\,(1-\mu)}{r_{1}}+\frac{2\,\mu}{r_{2}},
\end{equation}
where $\mu<1$ is the  mass ratio of the secondary and the primary, $r_1^2=(x+\mu)^2+y^2$ and $r_2^2=(x-1+\mu)^2+y^2$. Orbits therefore lie on a 3D subspace $C(x,y,\dot{x},\dot{y})=C$ embedded in the 4D phase space.  Points of an orbit that intersects a given surface, e.g. $y=0$, in a given direction, e.g. $\dot{y}>0$, lie on a 2D surface of section  $(x,\dot{x})$. An order $k$ resonance corresponds to a set of k islands on  the surface of section \citep{Winter&Murray1,Winter&Murray2}. Here, we prefer to define the surface of section by $\dot{x}=0$, allowing us to follow orbit intersections in the $(x,y)$-plane.

We choose a mass ratio $\mu=0.01$ that is small enough for perturbation theory to apply and Keplerian osculating elements to be used. These elements  $(a,e,\varpi)$  vary on a longer scale than the orbital period. 
We then set at $t=0$, $\dot{x}_0=0$, so that:
 \begin{equation}
 \label{doty}
 \dot{y}_0=\pm \sqrt{x_0^2+y_0^2+\frac{2\,(1-\mu)}{|x_0+\mu|}+\frac{2\,\mu}{|1-(x_0+\mu)|}-C} \ .
 \end{equation}
The transformation  from barycentric variables $(x_0,y_0, \dot{x}_0, \dot{y}_0)$ to astrocentric (centered on the primary) variables $(x_1,y_1, \dot{x}_1, \dot{y}_1)$ is given as:

\parbox{4cm}{
\begin{eqnarray*}
x_1 &=& x_0+\mu  \\
y_1 &=& y_0  
\end{eqnarray*}}
\hfill
\parbox{4cm}{
\begin{eqnarray}
\dot{x}_1 &=& \dot{x}_0-y_1   \\
\dot{y}_1 &=& \dot{y}_0+x_1 
\end{eqnarray}}

When $\dot{x}_0=0$ and $y_0=0$ we have $\dot{x}_1=0$ and $\dot{y}_1=\dot{y}_0+x_1$.  Hence, to set up prograde orbits we choose
$x_1>0$ and $\dot{y}_0>0$, or $x_1<0$ and $\dot{y}_0<0$. However,  to set up retrograde orbits we must have $x_1>0$ and $\dot{y}_0<-x_1$, or $x_1<0$ and $\dot{y}_0>-x_1$. By replacing $y_0=0$ and $\dot{y}_0=-x_1$ into Eq.~(\ref{doty}), we obtain a limit range on the Jacobi constant 
\begin{equation}
C< \frac{2\,(1-\mu)}{| x_0+\mu |}+\frac{2\,\mu}{| x_0-1+\mu |}-\mu (\mu+2\,x_0),
\end{equation}
such that, within this range of $C$,  choosing $x_1>0$ and $\dot{y}_0<0$, or $x_1<0$ and $\dot{y}_0>0$  ensures that the orbits are retrograde. When $|x_1|<3$, this limit is $C\lesssim 0.7$.
 
We construct  surfaces of section defined by  $\dot{x}=0$ and $\dot{y}\times\dot{y}_{0}>0$ so that the initial condition lies on the surface of section.  We vary  $x_0$ between -3 and 3 with increments of 0.05, and $y_0$ between 0 and 0.8 with increments of 0.1. The Jacobi constant in the range $-1.9\leq C\leq 0.7$ is incremented by 0.1.  We use Eq.~(\ref{doty}) to obtain $\dot{y}_0$ and  choose $\dot{y}_0>0$ if $x_1<0$ and $\dot{y}_0<0$ if $x_1>0$.  This ensures that the initial conditions always correspond to retrograde orbits (as $C\lesssim0.7$). The equations of motion of the CR3BP were numerically integrated for up to $10^6$ binary periods using a Bulirsch-Stoer algorithm with accuracy $10^{-14}$.  A selection of the surfaces of section will be discussed in Sect.~4.3. The full set of surfaces of section can be seen as Online Resource~5.

\begin{figure*}
 \begin{center}
  \sidesubfloat[]{\includegraphics*[width=4cm]{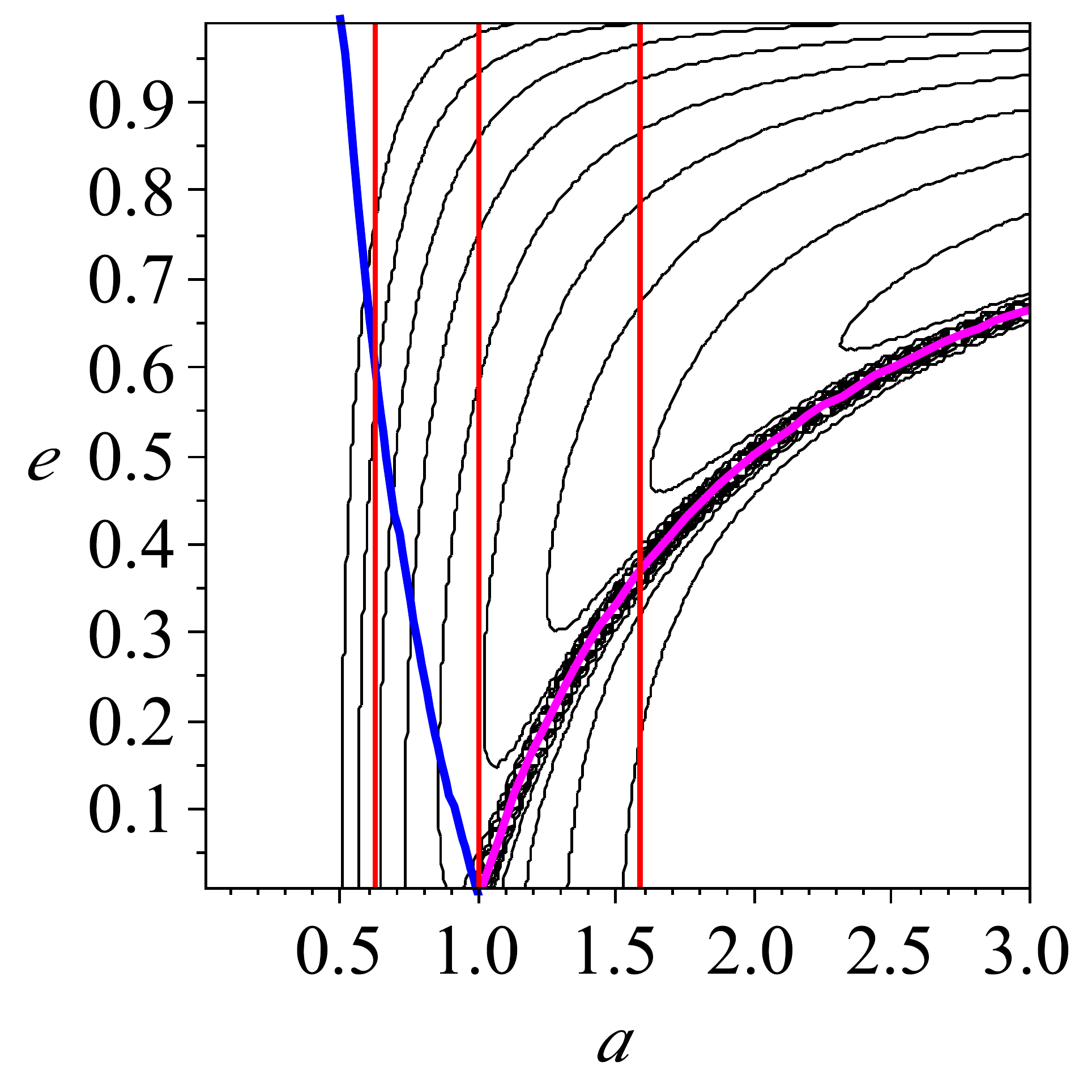}}\quad%
  \sidesubfloat[]{\includegraphics*[width=4cm]{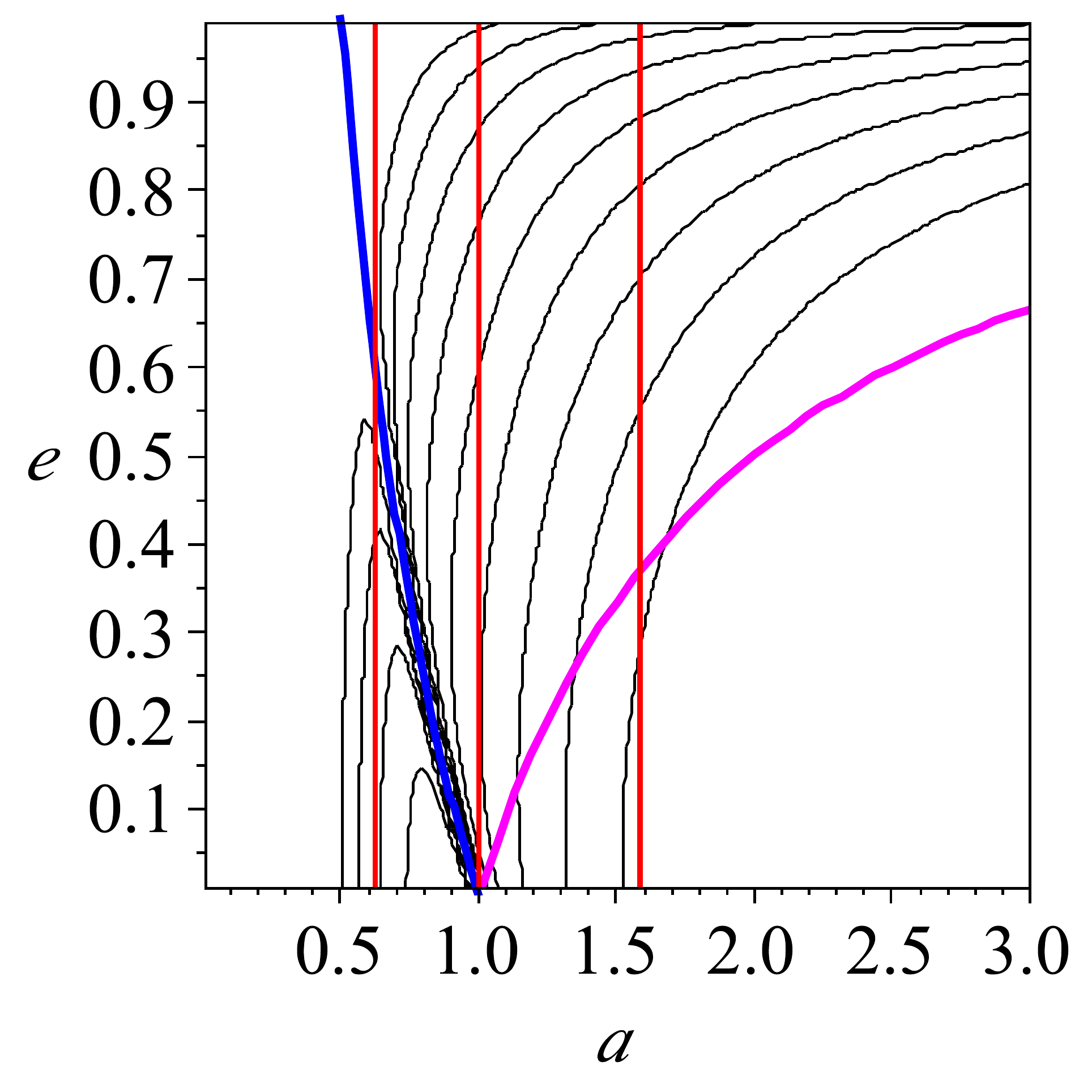}} \,
 \sidesubfloat[]{\includegraphics*[width=4cm]{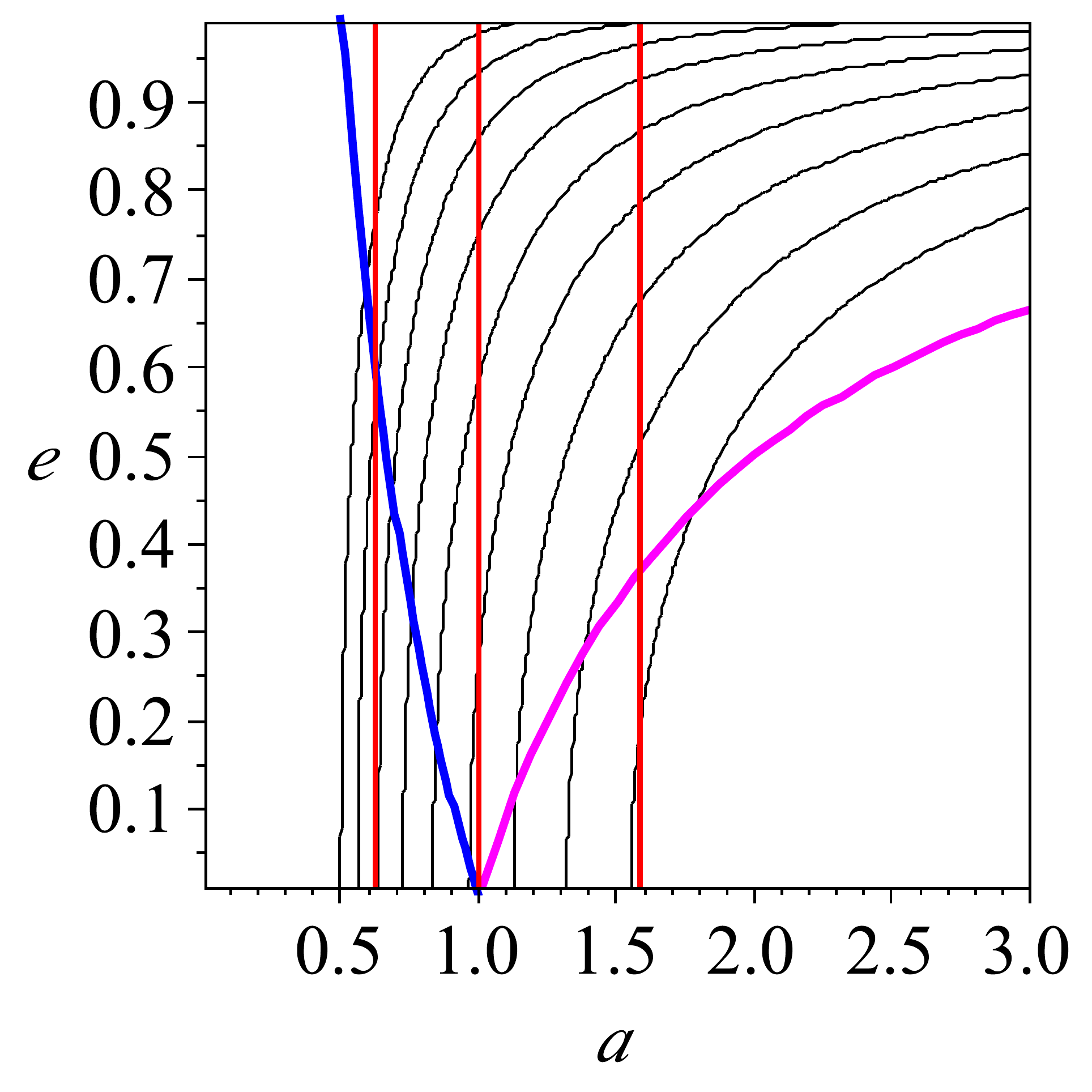}}\quad%
  \sidesubfloat[]{\includegraphics*[width=4cm]{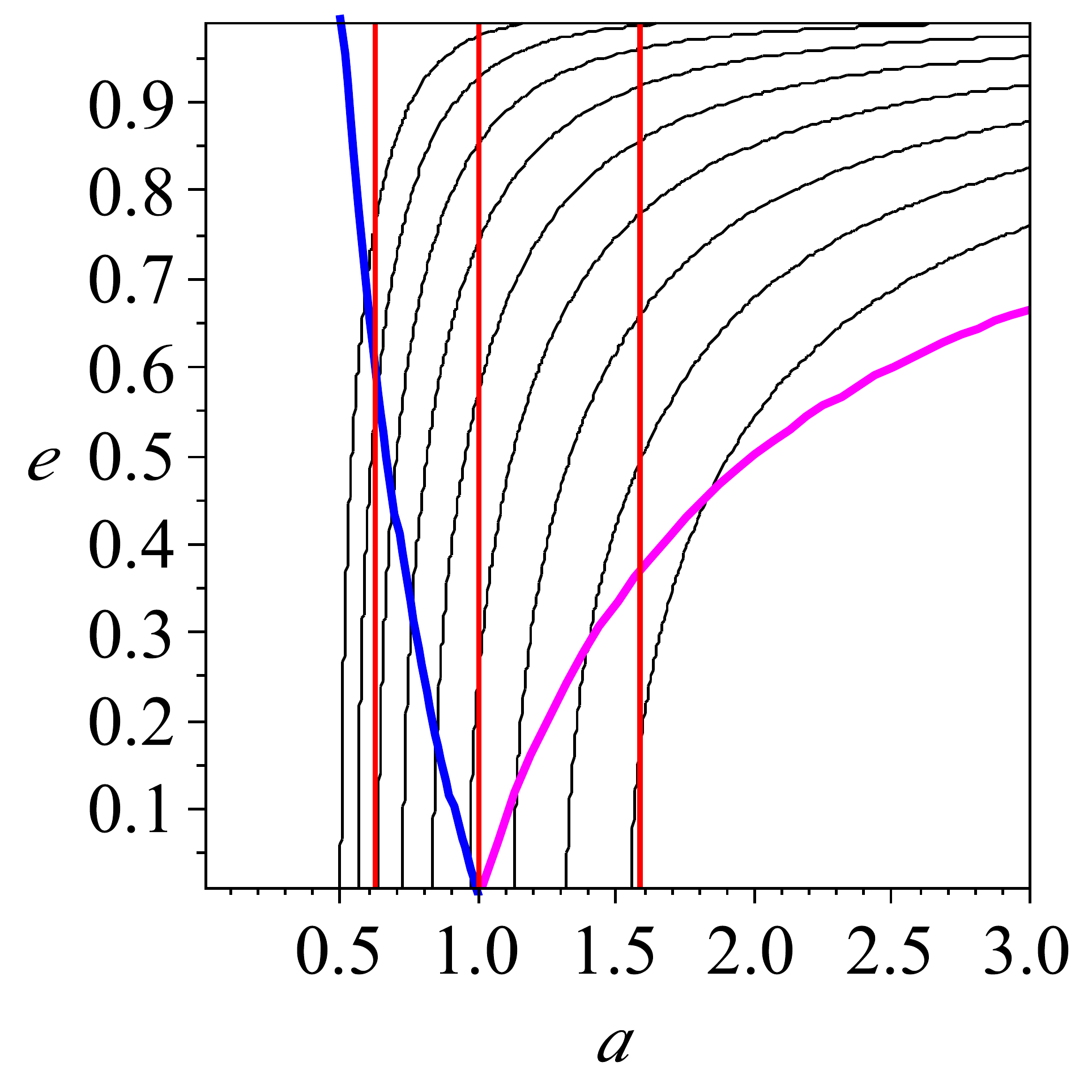}}  
\caption{Level curves of $C$ at values (from left to right) 0.6,0.3,0,-0.3,-0.6,-0.9,-1.2,-1.5,-1.8. Initial condition at conjunction and pericenter (a), conjunction and apocenter (b), opposition and pericenter (c), opposition and apocenter (d). The magenta and blue lines locate collision with secondary at pericenter or apocenter. The red vertical lines show location of 2/-1, 1/-1 and 1/-2 resonances.}    
\label{jacobilevels}
\end{center}
\end{figure*}

We show the level curves of constant $C$  in $(a,e)$ space for initial conditions at conjunction i.e. $x_1>0$ and $y=0$ (Fig.~\ref{jacobilevels} (a) and (b)) and for initial conditions at opposition i.e. $x_1<0$ and $y=0$  (Fig.~\ref{jacobilevels} (c) and (d)). The chosen range $-1.9\leq C\leq 0.7$ spans semi-major axes between 0.5 and 1.5, thus includes the 2/-1, 1/-1 and 1/-2 resonance regions.  

\subsection{Resonant angles}
The chosen initial conditions are $\lambda=\lambda'=0$ (conjunction:  $y=0$ and $x+\mu>0$) or $\lambda=180^\circ$ and $\lambda'=0$ (opposition:  $y=0$ and $x+\mu<0$). The points on the surface of section ($\dot{x}=0$) with $y=0$ correspond to the osculating orbits' pericenter  ($\lambda=\varpi$) or apocenter ($\lambda=\varpi+180^\circ$), depending on the $x$ and $C$ values. 

Starting in conjunction and at pericenter (apocenter) corresponds to the resonant angle
$\phi=q\,\lambda-p\,\lambda'-(p+q)\,\varpi=0 $ ($[p+q]\times 180^\circ$). In this case,  $\phi=0$ ($180^\circ$) if $p+q$ is even (odd). Starting at opposition and at pericenter (apocenter), $\phi=-p\times 180^\circ$ ($q\times 180^\circ $). Hence for opposition at pericenter, $\phi=0$ ($180^\circ$)  if $p$ is even (odd) and at apocenter $\phi=0$ ($180^\circ$)  if $q$ is even (odd). Therefore, even order resonant angles (such as that of the 1/-1 resonance)  may librate around 0 for initial conditions at conjunction, or around $180^\circ$ for initial conditions at opposition. Odd order resonant angles with $p$ even (such as that of the 2/-1 resonance) may librate around  180$^\circ$ for initial conditions at apocenter, or around 0 for initial conditions at pericentre. Odd order resonant angles with an odd $p$  (such as that of the 1/-2 resonance) may librate around  180$^\circ$ for initial conditions at conjunction and apocenter, opposition and pericenter, or  around 0 for initial conditions at conjunction and pericentre, opposition and apocenter.

\subsection{Results of numerical integrations}

\subsubsection{Resonant configurations}
We examine a selection of orbits in the vicinity of the 2/-1, 1/-1 and 1/-2 resonances shown in the frame rotating with the binary.  These orbits are periodic when at exact resonance and quasi-periodic otherwise.

Fig.~(\ref{xy21}) shows 2/-1 resonant orbits. The top left panel shows an orbit with  $C=+0.6$ that starts at conjunction or opposition ($y=0$), and pericenter (mode A). The top right panel shows an orbit with  $C=+0.6$ that starts at conjunction or opposition ($y=0$), and  apocenter (mode B). Resonant libration occurs around 0 in mode A and  around $180^\circ$ in mode B (Online Resource~1).
The low left panel shows  an orbit  with  $C=+0.3$ that starts at  conjunction or opposition ($y=0$),  and pericenter (mode A). The low right panel shows an orbit  with  $C=+0.3$ that starts at  conjunction or opposition ($y=0$), and apocenter (mode B). The latter orbit is very close to collision with the secondary.  

Fig.~(\ref{xy11}) shows  1/-1 resonant orbits. The top left panel shows an orbit  with  $C=+0.6$  that starts at $x_1>0$ and $y=0$, or  $x_1<0$ and $y\neq 0$ (mode I). Resonant libration occurs around 0 and disruptive close encounters are avoided despite the high eccentricity (Online Resource~2 left: mode I). The top right panel shows an orbit  with  $C=+0.6$ that starts at $x_1<0$ and $y=0$ (mode II).  The 1/-1 resonant angle librates around $180^\circ$  (Online Resource~2 right: mode II) but the orbit is close to collision and becomes unstable when $C<0.6$. Both orbits are described by the equilibria of the ponderomotive potential $S$ in Fig.~ \ref{figfat1} (2nd row, rightmost panel).

\begin{figure*}
\begin{center}
\includegraphics*[width=6cm]{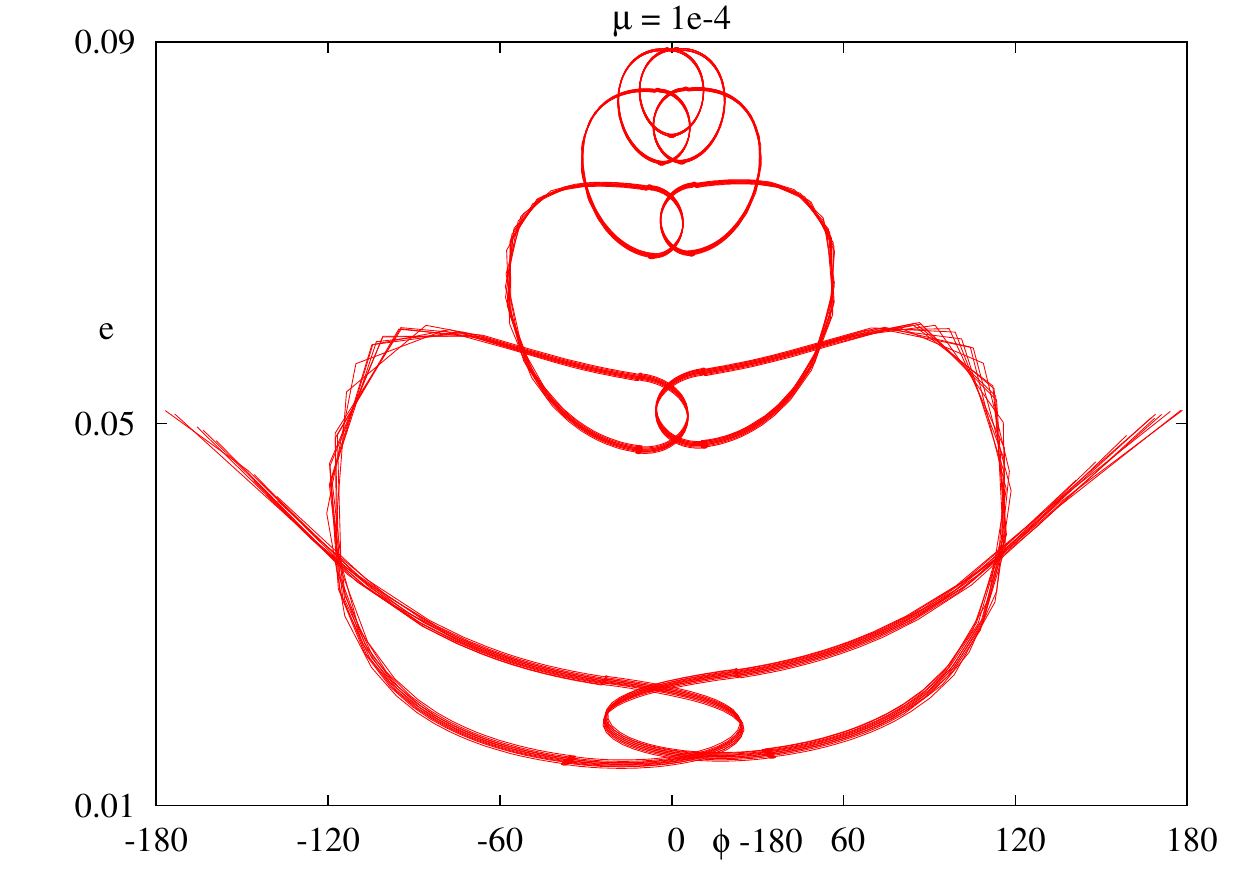}\includegraphics*[width=6cm]{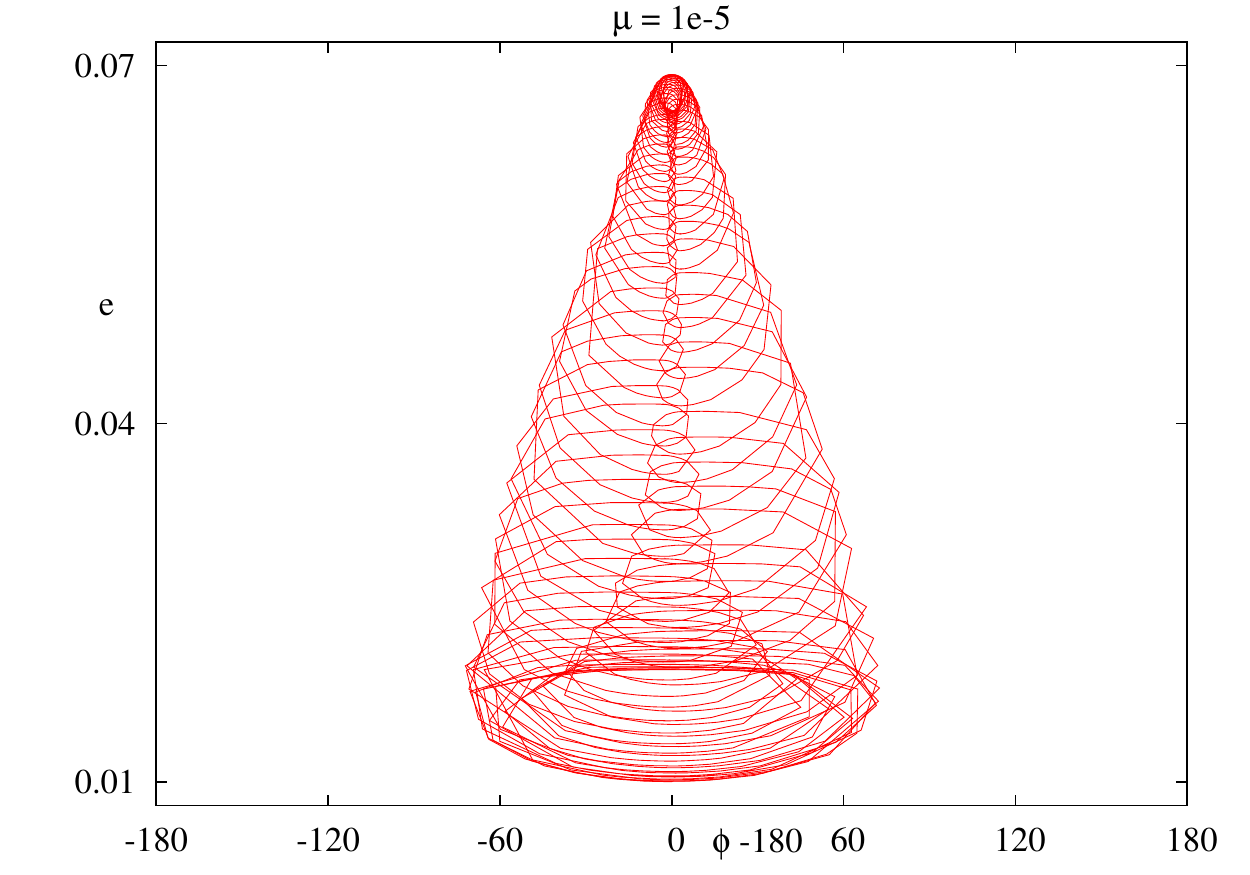} \\ \caption{Co-orbital resonant libration at small eccentricity. The orbits' initial conditions are $M=180^\circ$, $\varpi=0$ and for $\mu=10^{-5}$, $a/a^\prime-1=0.001$, $e=0.01$, whereas for $\mu=10^{-4}$ , $a/a^\prime-1=0.01$, $e=0.08$. For better visibility, orbits are shown only for 100 periods.}    
\label{figfat2}
\end{center}
\end{figure*}

The mid left panel of Fig.~(\ref{xy11}) shows an orbit  with  $C=-0.9$ and moderate eccentricity that starts at $x_1>0$ and $y=0$, or $x_1<0$ and $y\neq 0$ (mode I). The 1/-1 resonant angle $\phi$ librates around 0.  The mid  right panel  of Fig.~(\ref{xy11}) shows  an orbit  with  $C=-1.1$ that  is nearly circular  and has initially $x_1<0$  and $y=0$  (mode III).  This crossing orbit is very close to collision and we expect resonant libration  around $180^\circ$ (Online Resource~3 right: mode III). What happens is an interesting behavior best seen if we integrate a similar orbit for smaller mass ratios thus reducing the jitter due to close encounters. In Fig. (\ref{figfat2}, left panel), we plot in the $(\phi,e)$-plane, a similar orbit but with $\mu=10^{-4}$. The resonant argument alternates periodically between libration and circulation in a state that is stable over long time scales. Observing libration around $\phi=180^\circ$ requires a finer search which becomes easier  as the mass ratio is decreased (Fig. \ref{figfat2}, right panel).
The low left panel of  Fig.~(\ref{xy11}) shows an orbit  with  $C=-1.2$ that has small eccentricity and starts at $x_1>0$ and $y=0$,  or $x_1<0$ and $y\neq 0$ (mode I). This crossing orbit is also  close to collision and the resonant angle $\phi$ librates around 0 (Online Resource~3 left: mode I).  The low  right panel of Fig.~(\ref{xy11}) shows an orbit  with  $C=-1.2$ that is nearly circular and starts at  $x_1>0$ or $x_1<0$,  and $y=0$  for  $C=-1.2$. This is a non-crossing orbit just exterior to the secondary's orbit and the resonant angle circulates. 

Fig.~(\ref{xy12}) shows  1/-2 resonant orbits. The top left panel shows an orbit  with  $C=-1.5$ that  starts at conjunction and pericenter,  or opposition and apocenter (mode A). The top right panel shows  an orbit  with  $C=-1.5$ that  starts at conjunction and  apocenter,  or opposition and pericenter (mode B).  Resonant libration occurs around 0 in mode A and  around $180^\circ$ in mode B (Online Resource~4). When $C=-1.5$ both mode A and mode B orbits are close to collision with the secondary.
The low left panel shows an orbit  with  $C=-1.2$ that  starts at conjunction and pericenter, or opposition and apocenter (mode A). This is the only stable configuration when $C>-1.3$. The low right panel shows an orbit with $C=-1.8$  that  starts  at conjunction and apocenter,   or opposition and pericenter (mode B). This is the only stable configuration when $C=-1.8$.

\subsubsection{Surfaces of section}
\begin{figure*}
 \begin{center}
    \includegraphics*[width=6.2cm]{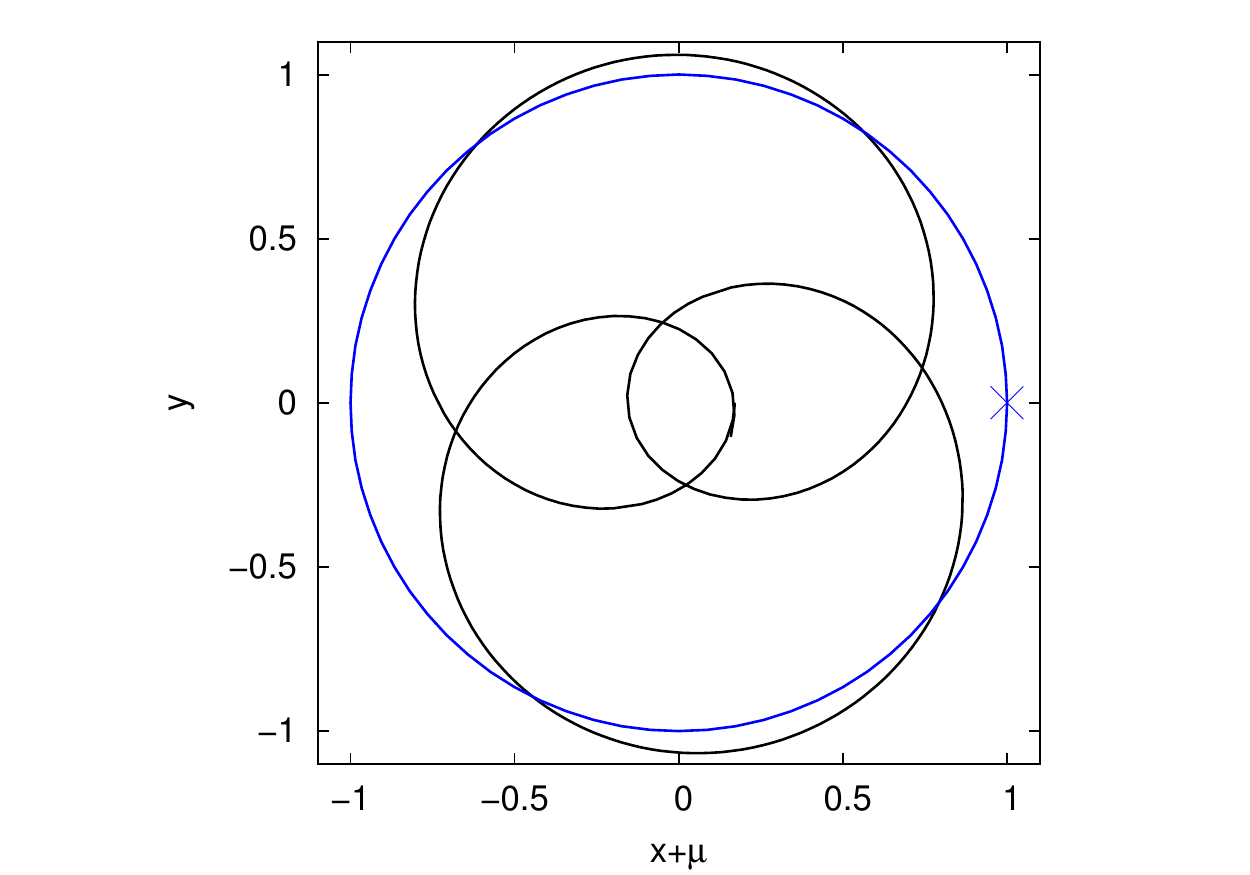}\includegraphics*[width=6.2cm]{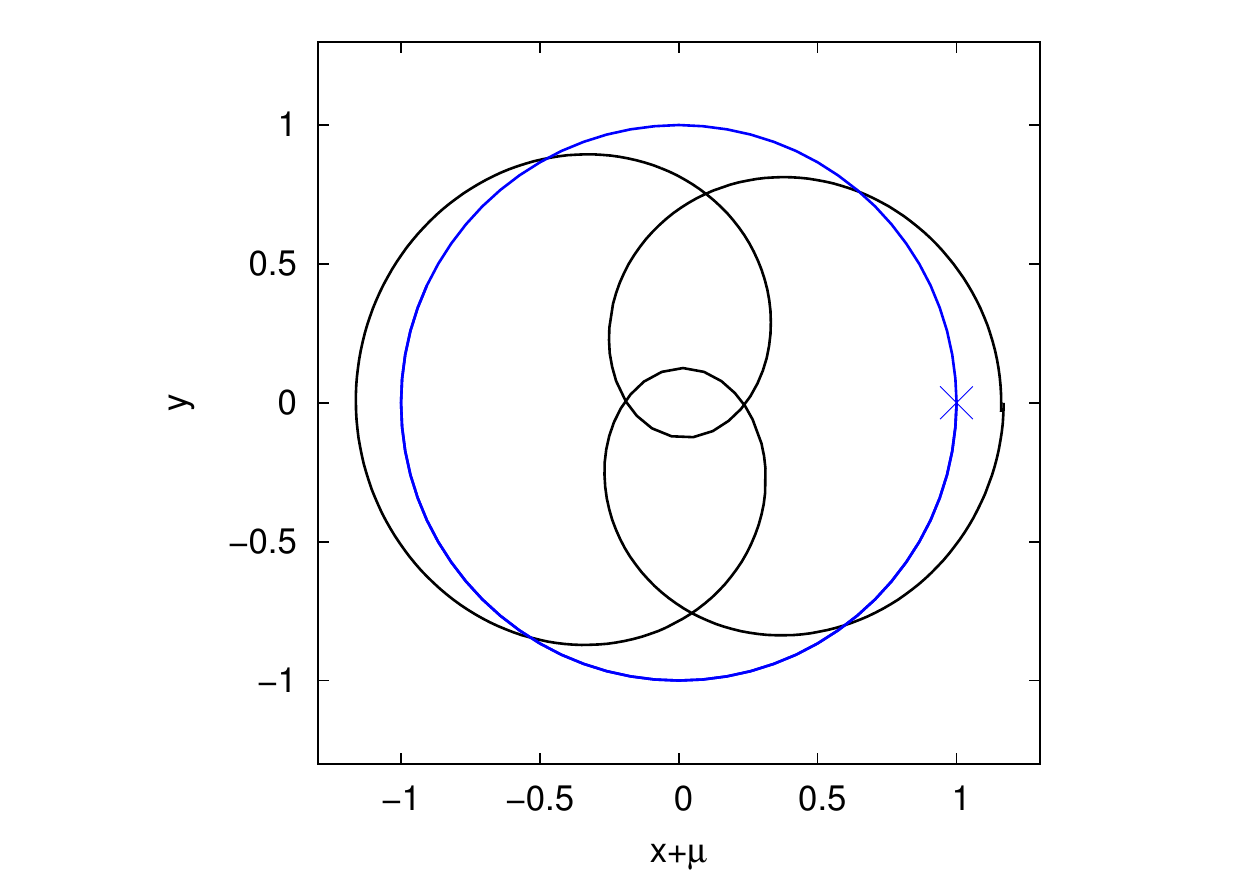}  \\
     \includegraphics*[width=6.2cm]{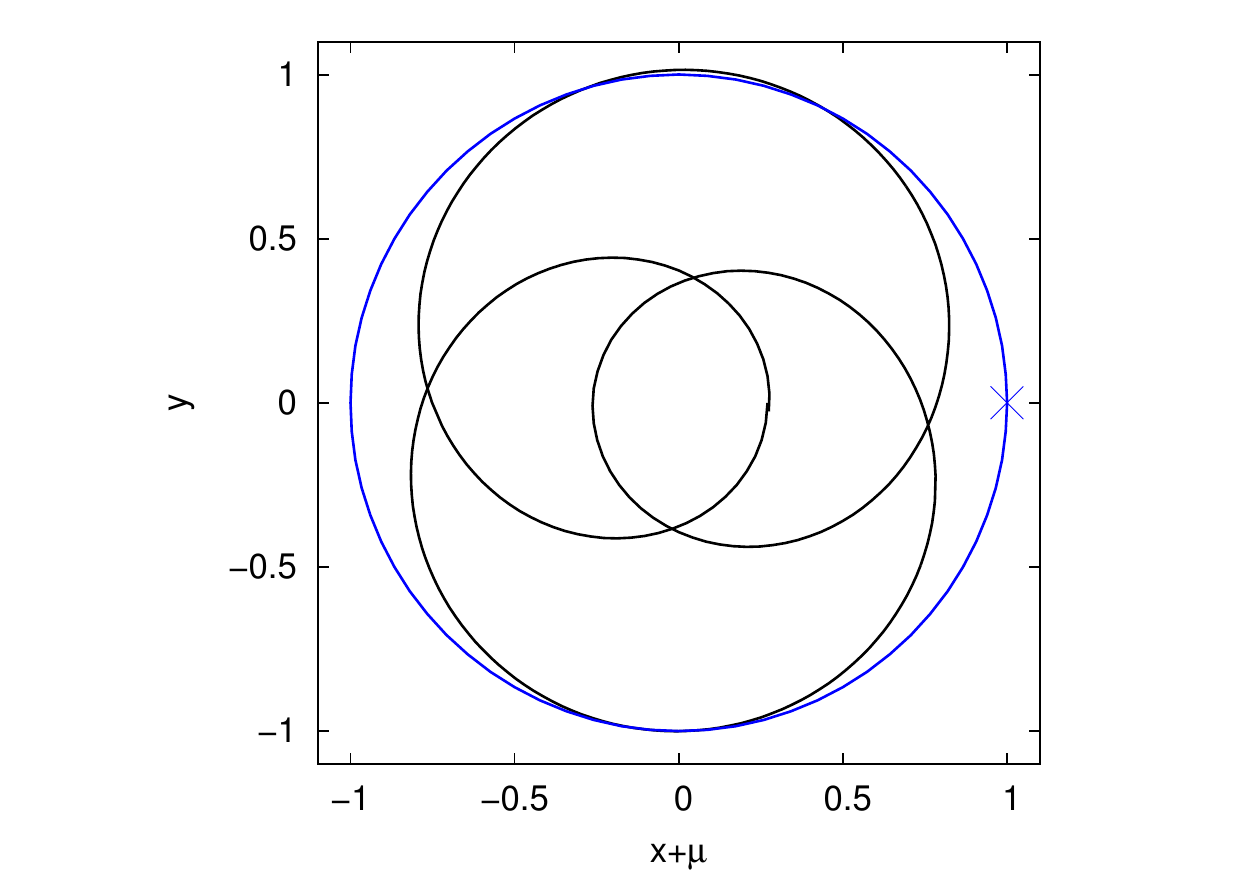}\includegraphics*[width=6.2cm]{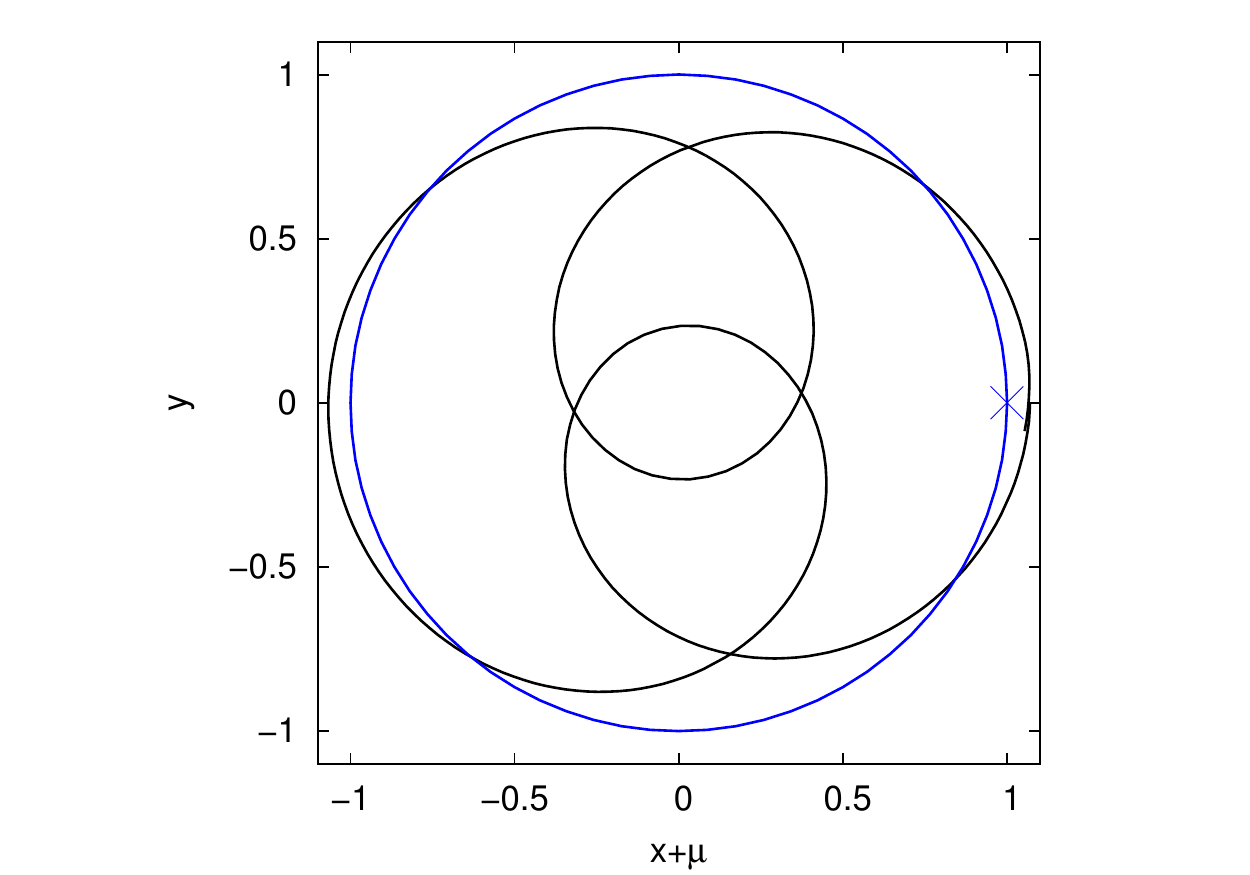}  \\   
\caption{Orbits in 2/-1 resonance seen in synodic frame: $C=0.6$ mode A (top left) and mode B (top right);
$C=0.3$ mode A (low left) and mode B (low right). A unit radius circle in blue helps identify the crossing orbits and  non-crossing orbits.}    
\label{xy21}
\end{center}
\end{figure*}

\begin{figure*}
 \begin{center}
    \includegraphics*[width=6.2cm]{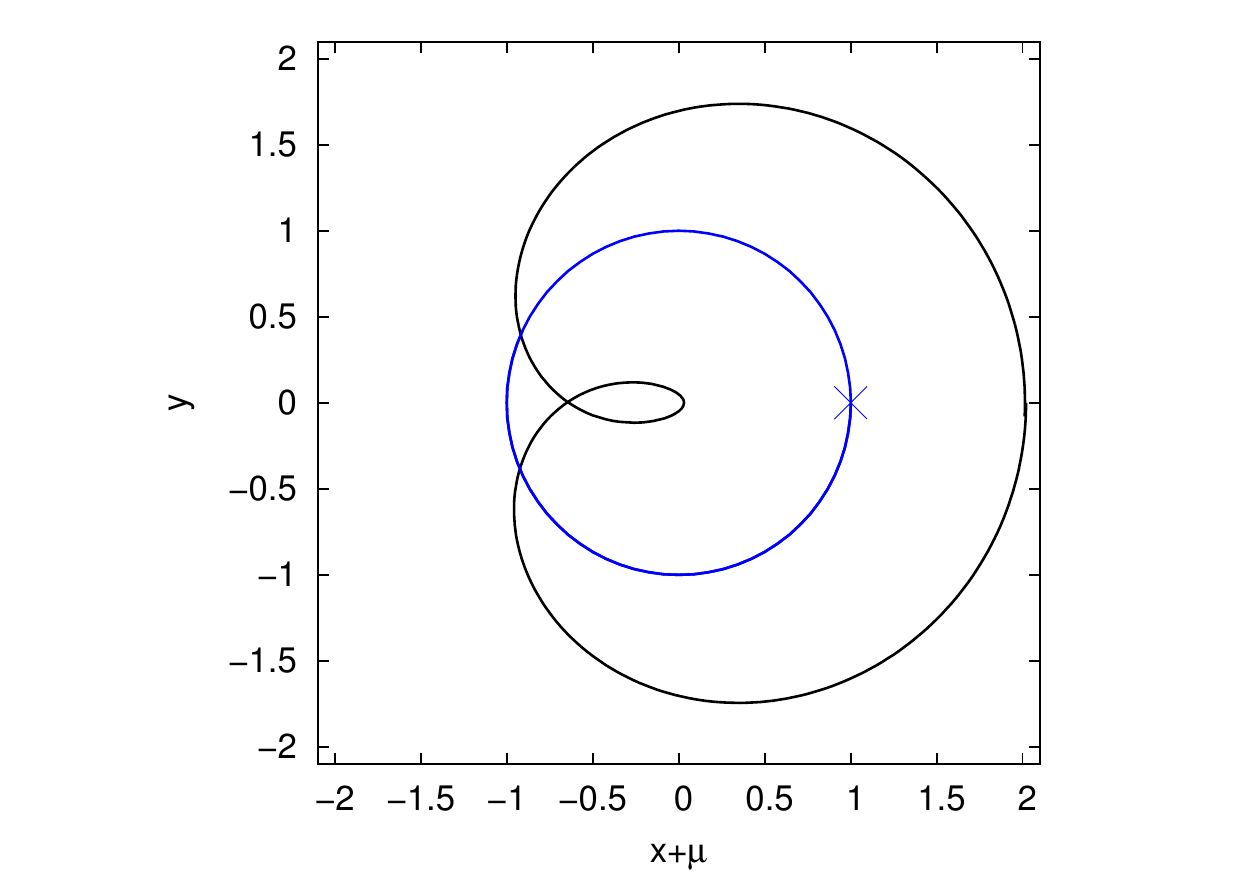}\includegraphics*[width=6.2cm]{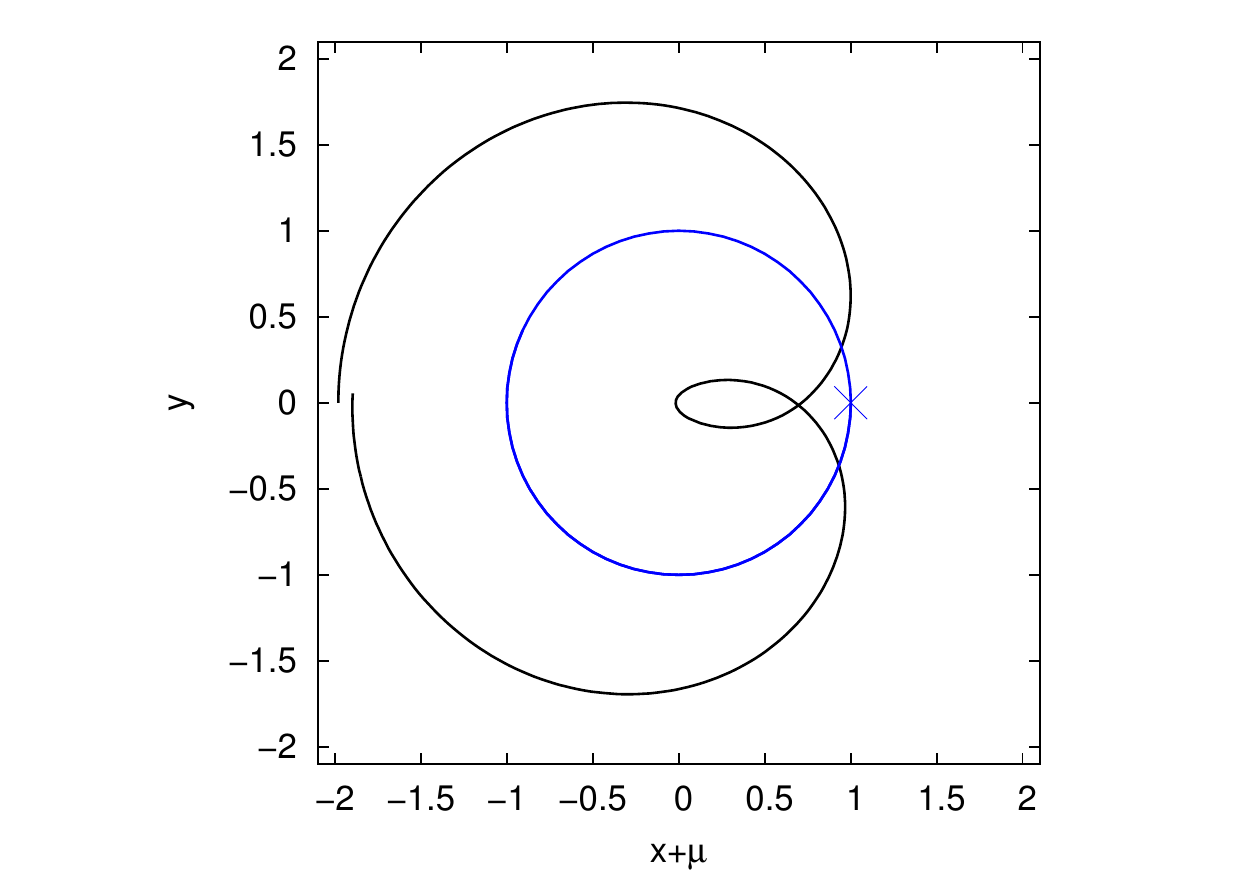}  \\
    \includegraphics*[width=6.2cm]{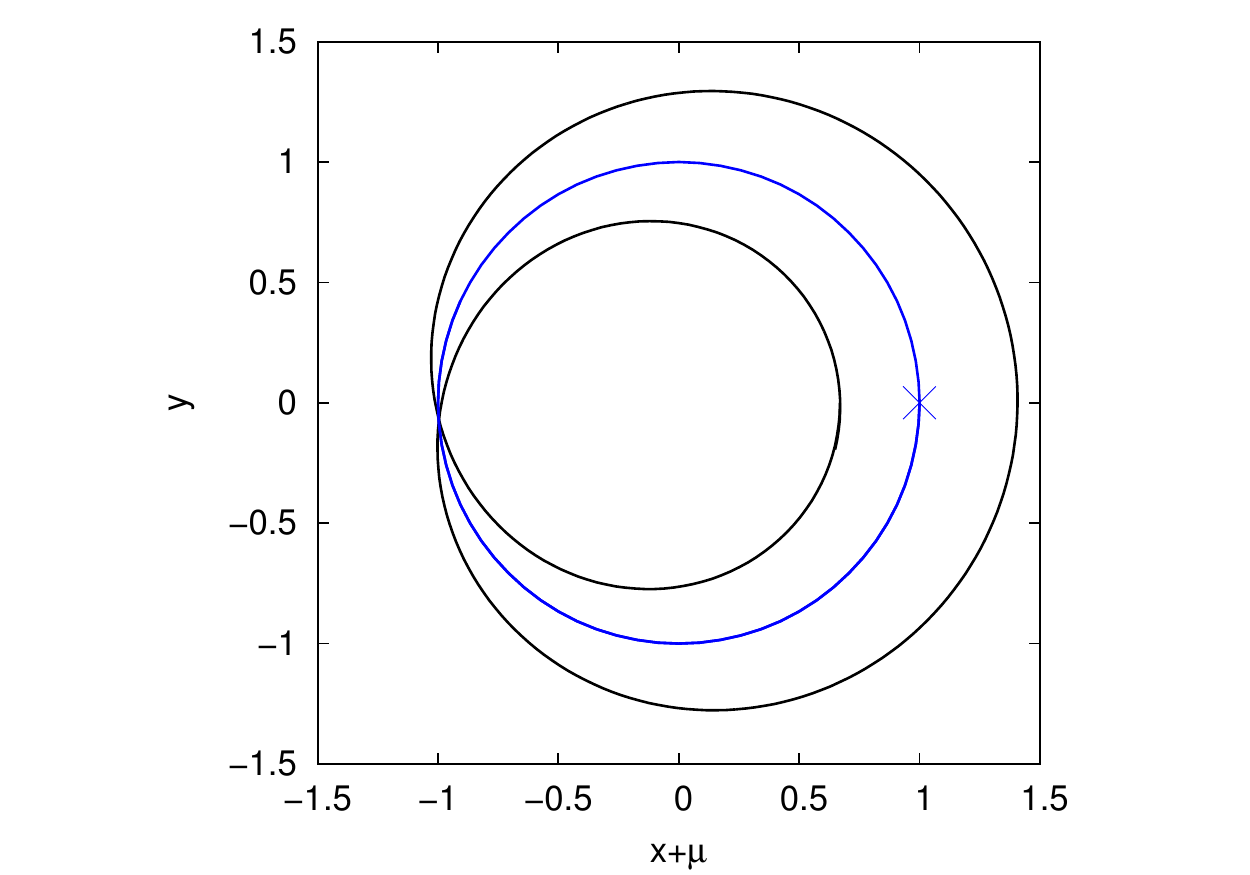}\includegraphics*[width=6.2cm]{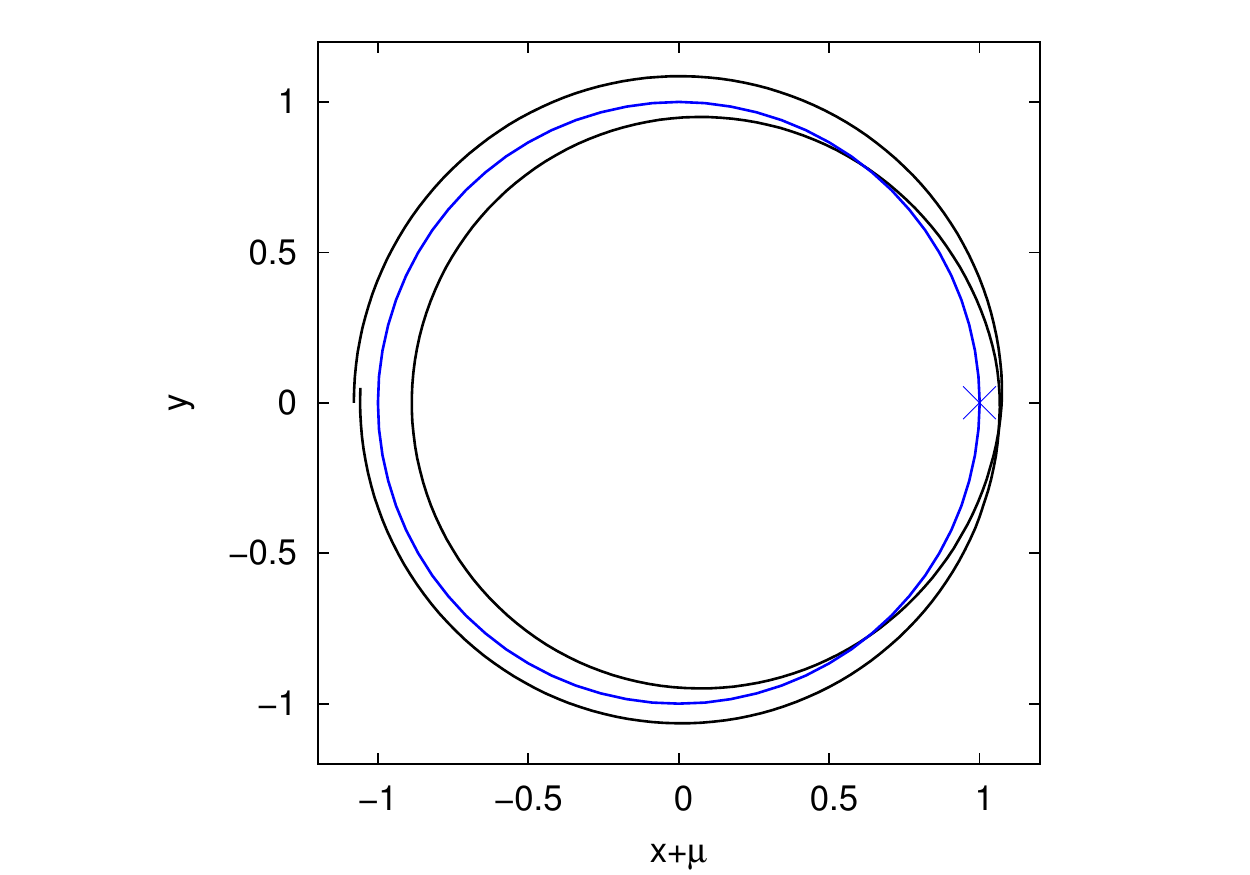}  \\
    \includegraphics*[width=6.2cm]{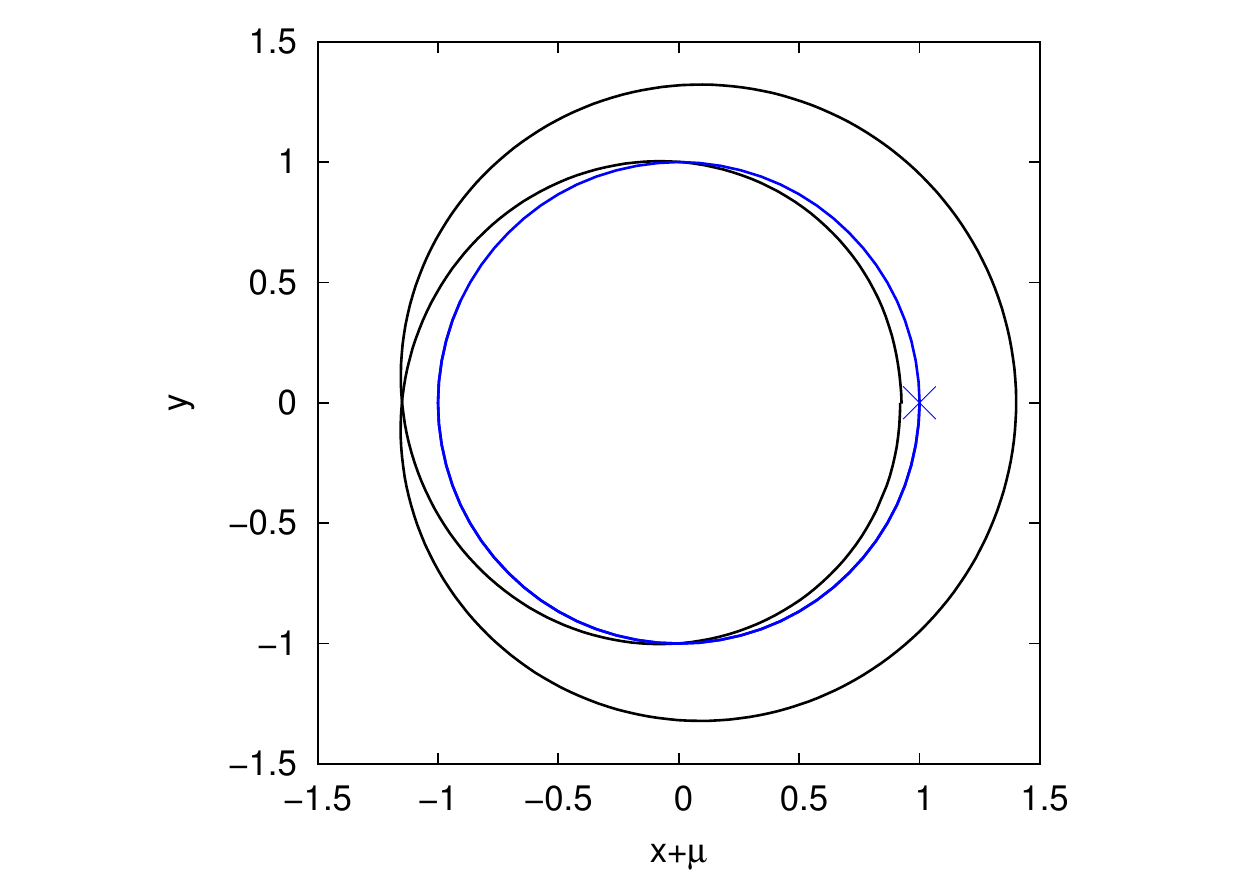}\includegraphics*[width=6.2cm]{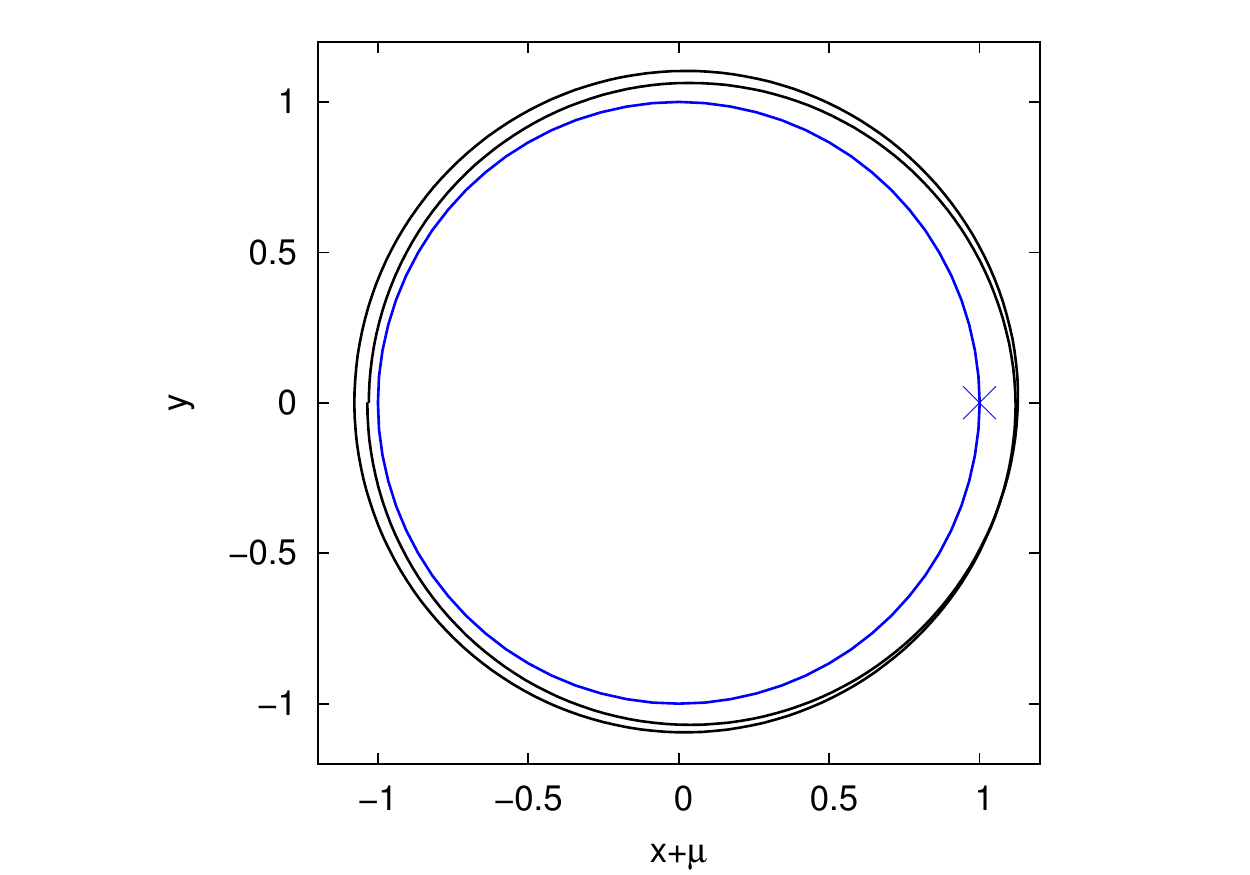}   \\ 
\caption{Orbits in 1/-1 resonance seen in synodic frame: $C=0.6$ mode I (top left) and mode II (top right); 
$C=-0.9$ mode I (mid left); $C=-1.1$ mode III libration (mid right);
$C=-1.2$ mode I (low left) and mode III circulation (low right).  A unit radius circle in blue helps identify the crossing orbits and  non-crossing orbits. }  
\label{xy11}    
\end{center}
\end{figure*}

\begin{figure*}
 \begin{center}
    \includegraphics*[width=6.2cm]{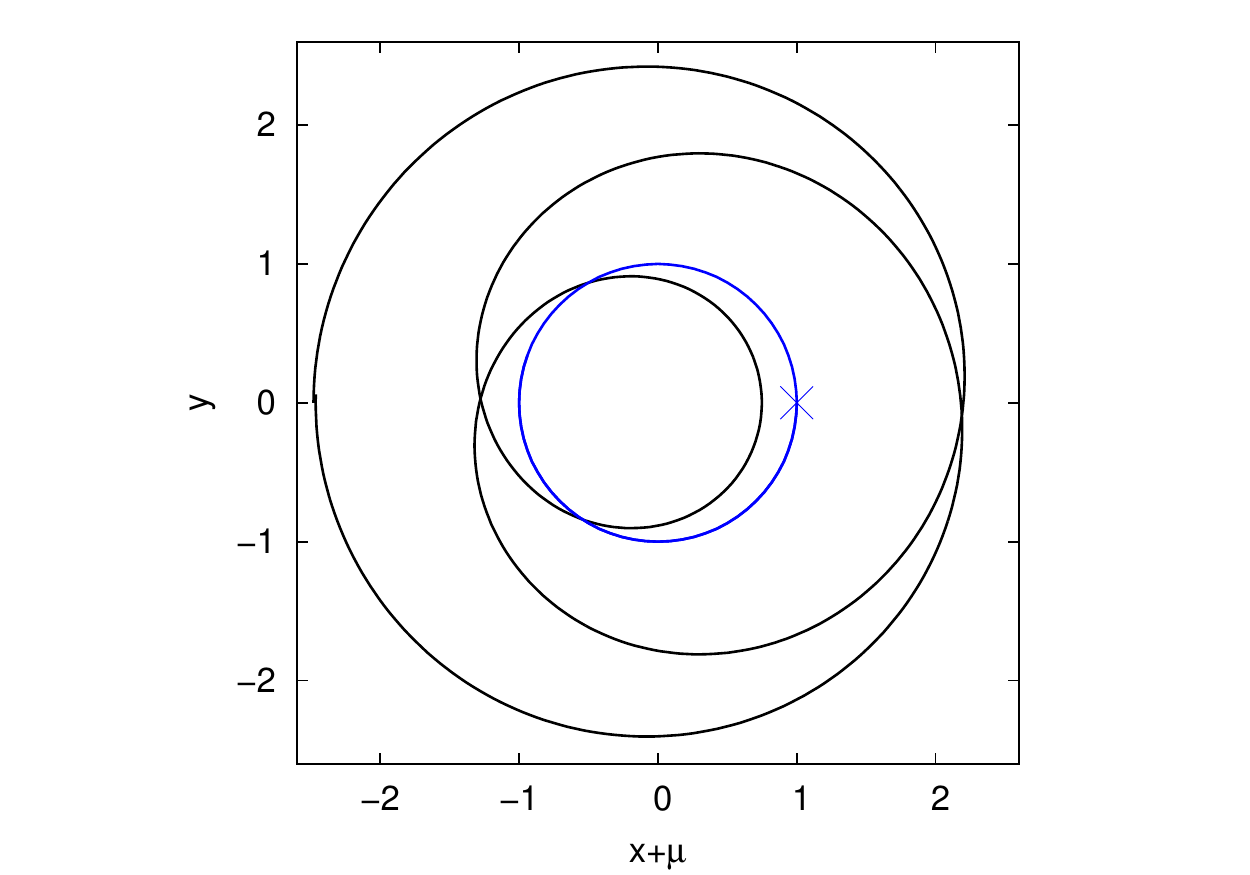}\includegraphics*[width=6.2cm]{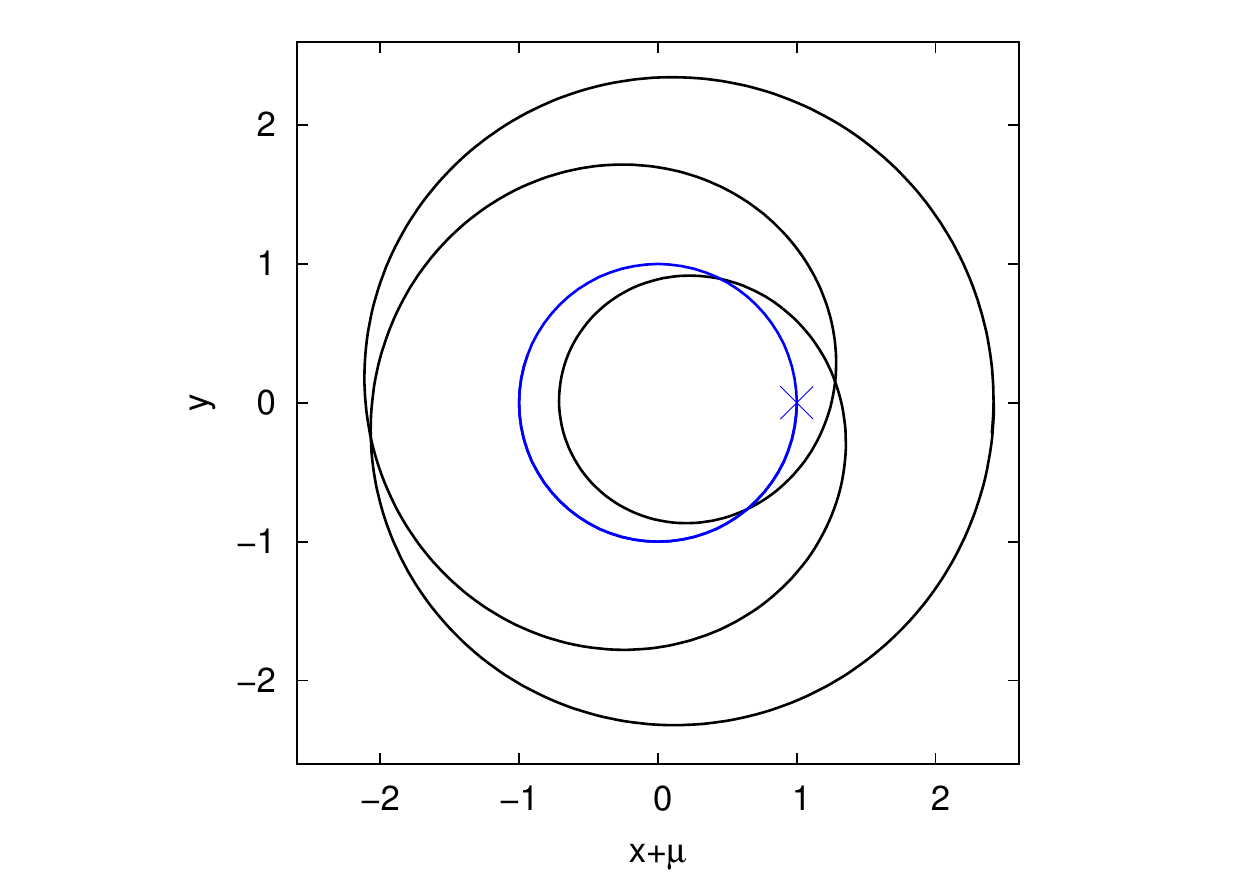}  \\
     \includegraphics*[width=6.2cm]{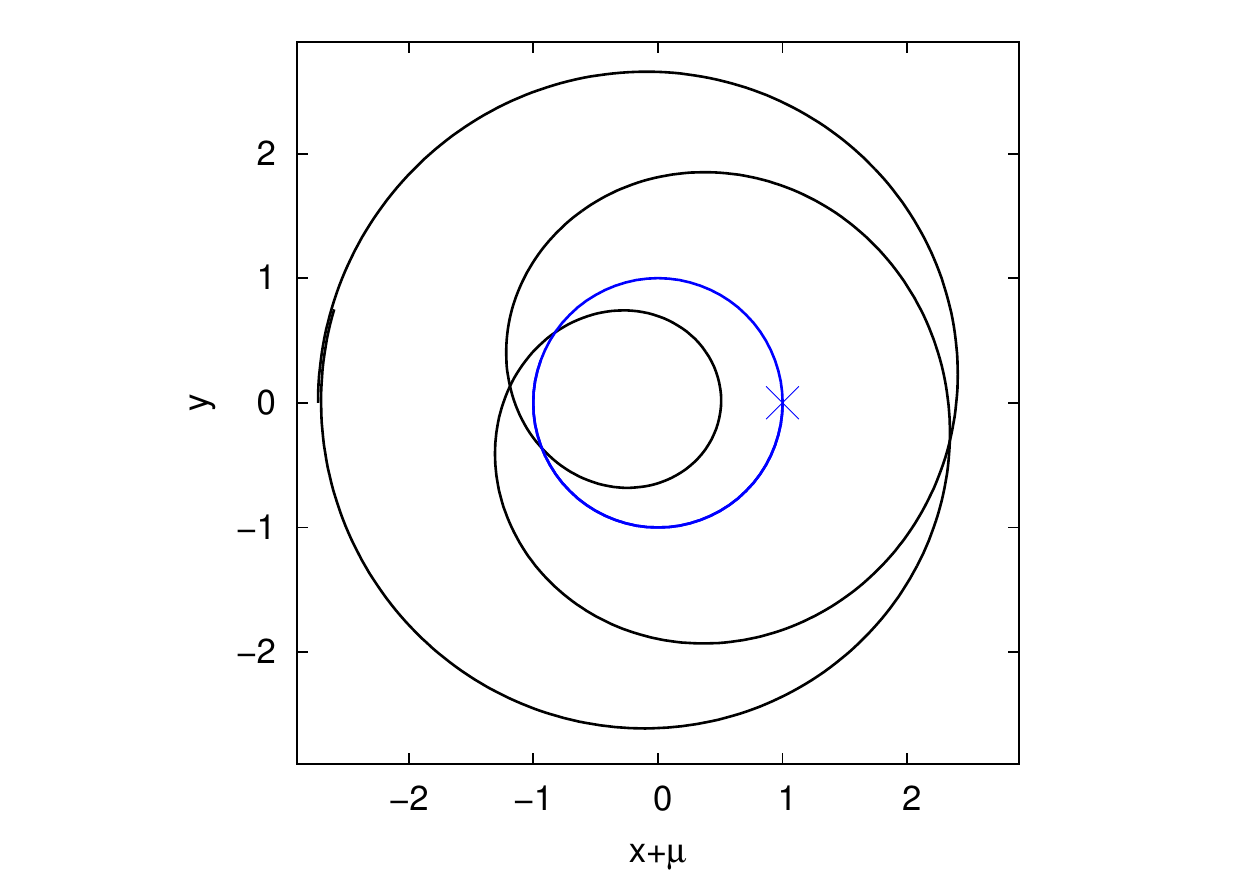}\includegraphics*[width=6.2cm]{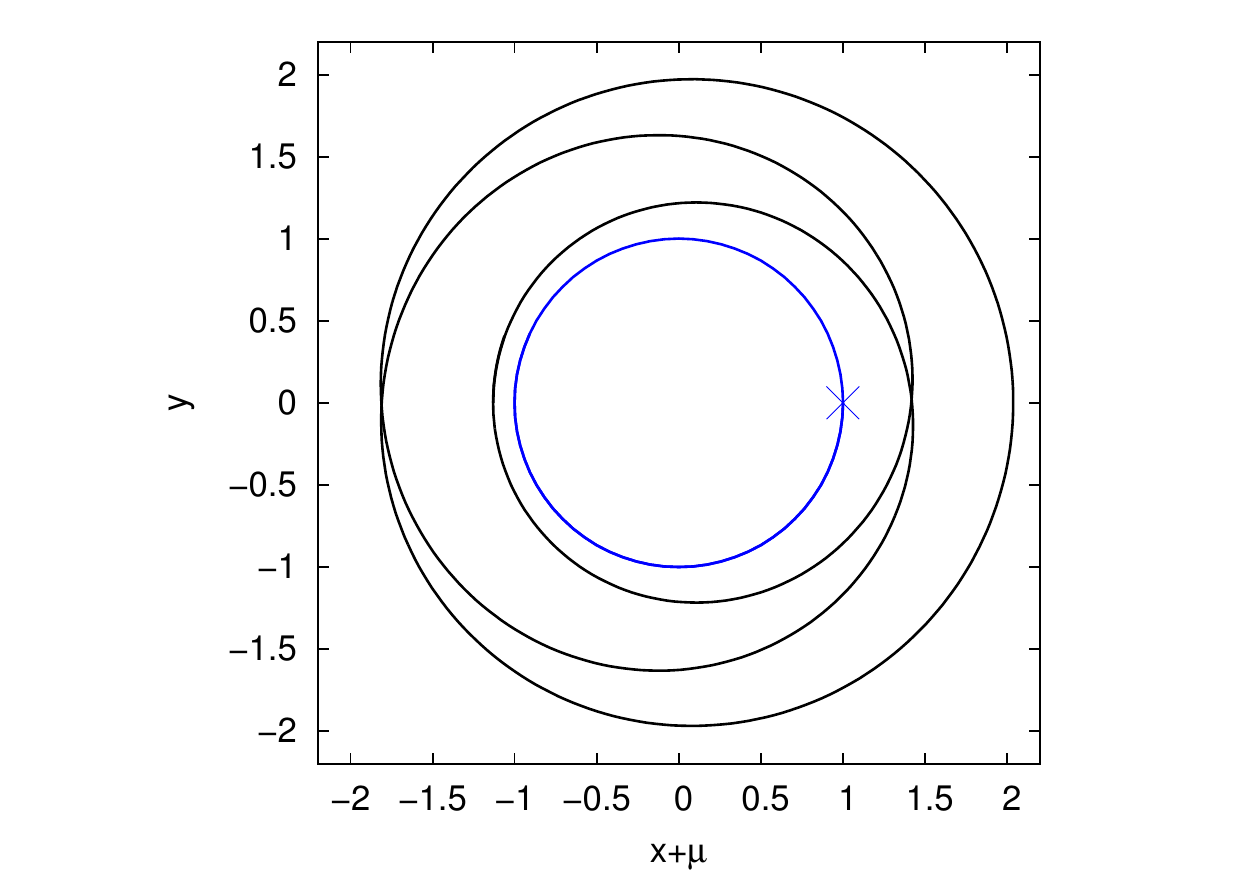}  \\   
\caption{Orbits in 1/-2 resonance seen in synodic frame: $C=-1.5$ mode A (top left) and mode B (top right);
$C=-1.2$ mode A (low left) and $C=-1.8$ mode B (low right). A unit radius circle in blue helps identify the crossing orbits and  non-crossing orbits.}   
\label{xy12} 
\end{center}
\end{figure*}

As seen in Figs.~\ref{xy21},~\ref{xy11},~\ref{xy12},  an order $k$ resonant orbit intersects the section $\dot{x}=0$ at  $2k$ different points. However, owing to the constraint on the sign of $\dot{y}$, we only see $k$ of these intersections on the surface of section $(x_1,y)$. Therefore,  in the surfaces of section (Fig.~\ref{sections}) a set of $k$ intersections on the left hand side ($x_1<0$) usually represents the same configuration (in the synodic frame) as a set of $k$ intersections on the right hand side ($x_1>0$). 

Fig.~\ref{sections} shows a selection of surfaces of section with $C=0.6$, $C=0.3$, $C=0.0$, $C=-0.9$, $C=-1.1$, $C=-1.2$, $C=-1.5$ and $C=-1.8$.  The full set of surfaces of section ($C$ between $0.7$ and $-1.9$ at steps 0.1) can be seen as Online Resource~5.  At these values of $C$ many initial conditions correspond to  crossing orbits (see Fig.~\ref{jacobilevels}) which can only be stable in resonance. Non-crossing small eccentricity orbits exist in the regions marked in green  (left and right on Fig.~\ref{sections}). Other colors correspond to different libration modes as described below. Empty areas in the surfaces of section correspond to  initial conditions that lead to collision or escape. 

When $C=0.6$, we see nearly circular non-crossing orbits in the vicinity of the  3/-1 resonance (green) for initial conditions at opposition (left) or conjunction (right). Collision with the secondary occurs between the 3/-1 and 2/-1 resonances.  
We see islands of libration in the  2/-1 resonance. The 2/-1 resonant orbits  that start at pericenter (magenta on left and right)  correspond to the same configuration in the synodic frame (Fig.~\ref{xy21} top left panel: mode A) where the resonant angle librates around 0 (Online Resource~1 left). The 2/-1 resonant orbits that start at apocenter  (blue on left and right) also correspond to the same configuration in the synodic frame (Fig.~\ref{xy21} top right panel: mode B) where the resonant angle librates around $180^\circ$ (Online Resource~1 right). 

When $C=0.6$, we also see islands of libration  in the 1/-1 resonance that correspond to orbits with very high eccentricity values.  The 1/-1 resonant orbits that start at conjunction (black on right) or with $x_1<0$ and $y\neq 0$ (black on left) correspond to the same configuration in the synodic frame (Fig.~\ref{xy11} top left: mode I) where the resonant angle librates around 0 (Online Resource~2 left). The 1/-1 resonant orbit that starts  at opposition (red) is close to collision (Fig.~ \ref{xy11} top right: mode II) and the resonant angle librates around $180^\circ$ (Online Resource~2 right).   

When $C=0.3$, we see orbits the vicinity of the  2/-1 resonance for initial conditions at opposition (left) or conjunction (right).  The 2/-1 resonant angle can  circulate  for non-crossing orbits (green on left and right),  it can librate in mode A (Fig.~\ref{xy21} low left) for initial conditions at pericenter (magenta on left and right), or it can librate in mode B (Fig.~\ref{xy21} low right) for initial conditions at apocenter (blue on left and right).
Collision with the secondary  occurs  just outside the 2/-1 resonance separatrix. The 1/-1 resonant angle librates around 0 (mode I) for initial conditions at conjunction (black on right) or with $x_1<0$ and $y\neq 0$  (black on left). There are also islands of libration in the 1/-2 resonance for initial conditions at conjunction / pericenter (magenta on right) or opposition / apocenter (magenta on left). These correspond to the same configuration in the synodic frame where the resonant angle librates around 0 (Fig.~\ref{xy12} mode A).

When $C=0.0$, we see nearly circular orbits in the vicinity of the  2/-1 resonance (green) for initial conditions at opposition (left) or conjunction (right).  The 2/-1 resonant angle can only circulate and all these orbits are non-crossing.  Collision with the secondary occurs between the 2/-1 and 3/-2 resonances.   The 1/-1 resonant angle librates around 0 (mode I) for initial conditions at conjunction (black on right) or with $x_1<0$ and $y\neq 0$  (black on left).  The 1/-2 resonant angle librates around 0 (mode A) for  initial conditions at conjunction / pericenter (magenta on right) or opposition /apocenter (magenta on left).
 
When $C=-0.9$, all initial conditions correspond to crossing orbits hence they are only stable in resonance.  
The 1/-1 resonant angle librates around 0 (mode I) for initial conditions at conjunction (black on right) or with $x_1<0$ and $y\neq 0$  (black on left).  Collision with the secondary occurs in the 1/-1  resonance region.
The 1/-2 resonant angle librates around 0 (mode A) for  initial conditions at conjunction / pericenter (magenta on right) or opposition /apocenter (magenta on left).

When $C=-1.1$,  nearly circular orbits in the 1/-1 resonance starting at opposition (red on left) correspond to libration around $180^\circ$ (Fig.~\ref{xy11} mid right: mode III) although that is not very clear from the behavior of the resonant angle (Online Resource~3 right: mode III) due to the effect of repeated very close encounters. These are crossing orbits which  are very close to the collision boundary. The nearly circular orbit starting at conjunction (red on right) correspond to the 1/-1 resonance separatrix.  The 1/-1 resonant angle can also librate around 0  (mode I) for initial conditions at conjunction (black on left) or with $x_1<0$ and $y\neq 0$  (black on right). These are also crossing orbits close to the collision boundary.
The 1/-2 resonant angle librates around 0 (mode A) for  initial conditions at conjunction / pericenter (magenta on right) or opposition /apocenter (magenta on left).

When $C=-1.2$,  we see nearly circular orbits in the vicinity of the 1/-1 resonance (green)  for initial conditions at opposition (left) or conjunction (right). The 1/-1 resonant angle circulates and the orbits are just exterior to the secondary's orbit thus very close to the collision boundary (Fig.~\ref{xy11} low right: mode III).  
The 1/-1 resonant angle can also librate around 0 (Online Resource~3 left and Fig.~\ref{xy11} low left: mode I) for initial conditions at conjunction (black on right) or with $x_1<0$ and $y\neq 0$  (black on left). These are crossing orbits close to the collision boundary.  The 1/-1 resonance is no longer possible when $C=-1.3$ (see Online Resource~5).
When $C=-1.2$, the 1/-2 resonant angle librates around 0 (mode A) for  initial conditions at conjunction / pericenter (magenta on right) or opposition /apocenter (magenta on left). When $C=-1.3$  (see Online Resource~5) the 1/-2 resonant angle can also librate around $180^\circ$ (mode B) for  initial conditions at conjunction / apocenter  or  opposition / pericenter.

When $C=-1.5$, we see nearly circular orbits in the vicinity of the  2/-3 resonance (green) for initial conditions at opposition (left) or conjunction (right). Collision with the secondary occurs in the 2/-3 resonance region.
There are islands of libration in the 1/-2 resonance. The 1/-2 resonant orbits  that start at conjunction / pericenter  (magenta on right) and  opposition / apocenter (magenta on left)  correspond to the same configuration in the synodic frame (Fig.~\ref{xy12} top left panel: mode A) where the resonant angle librates around 0 (Online Resource~4 left). The 1/-2 resonant orbits  that start at conjunction / apocenter (blue on right) and opposition / pericenter (blue on left)  correspond to the same configuration in the synodic frame (Fig.~\ref{xy12} top right panel: mode B) where the resonant angle librates around  $180^\circ$ (Online Resource~4 right).   

When $C=-1.8$, there are nearly circular orbits in the vicinity of the  1/-2 resonance (green) for initial conditions at opposition (left) or conjunction (right).  The 1/-2 resonant angle can circulate or it can librate  around $180^\circ$ (mode B) for initial condition at opposition / pericenter (blue on left) or conjunction / apocenter (blue on right). Collision with the secondary occurs  just outside the 1/-2 resonance separatrix. The 1/-2 resonant angle can no longer librate around 0 (mode A).

\subsection{Analytic model for 2/-1 and 1/-2 resonances}
The structure of the 2/-1 and 1/-2 resonances, at low to moderate eccentricities, can be described by  the analytic model for 3rd order resonance presented in \citet{ssdbook} and described in Paper I. The Hamiltonian is:
\begin{equation}
\label{hamiltonian}
H=\frac{\delta}{2} (X^2+Y^2)+\frac{1}{4} (X^2+Y^2)^2\mp2\,X (X^2-3\,Y^2)
\end{equation}
where  $\delta$  measures the proximity to exact resonance, $\mp$ applies to 2/-1 or 1/-2 resonances, $X=R\cos(\phi/3)$, $Y=R\sin(\phi/3)$, and $R$ is a scaling factor specific for each resonance and dependent  on the resonance coefficient and on the mass ratio $\mu$ \citep{ssdbook}. For the 2/-1 resonance $\phi=\lambda-2\,\lambda^\prime-3\,\varpi$ and $R= 2^\frac{2}{3}\,\gamma\,e$, where $\gamma=(3\mu\,f_{82})^{-1}$ and
 $f_{82}=0.402$ is the amplitude of term 4D3.1 ($j=2$) when $\alpha=0.623$ . For the 1/-2 resonance $\phi=2\,\lambda-\lambda^\prime-3\,\varpi$ and $R= 2^2\,\gamma\,e$ where $\gamma=(3\mu\,(f_{85}-0.5/\alpha^2))^{-1}$ where 
 $f_{85}-0.5/\alpha^2=0.533$ is the combined amplitude of the terms 4D3.4 ($j=2$) and 4I3.6   when $\alpha=0.623$.
 
\begin{figure*}
\begin{center}
 \subfloat[]{\includegraphics*[width=6cm]{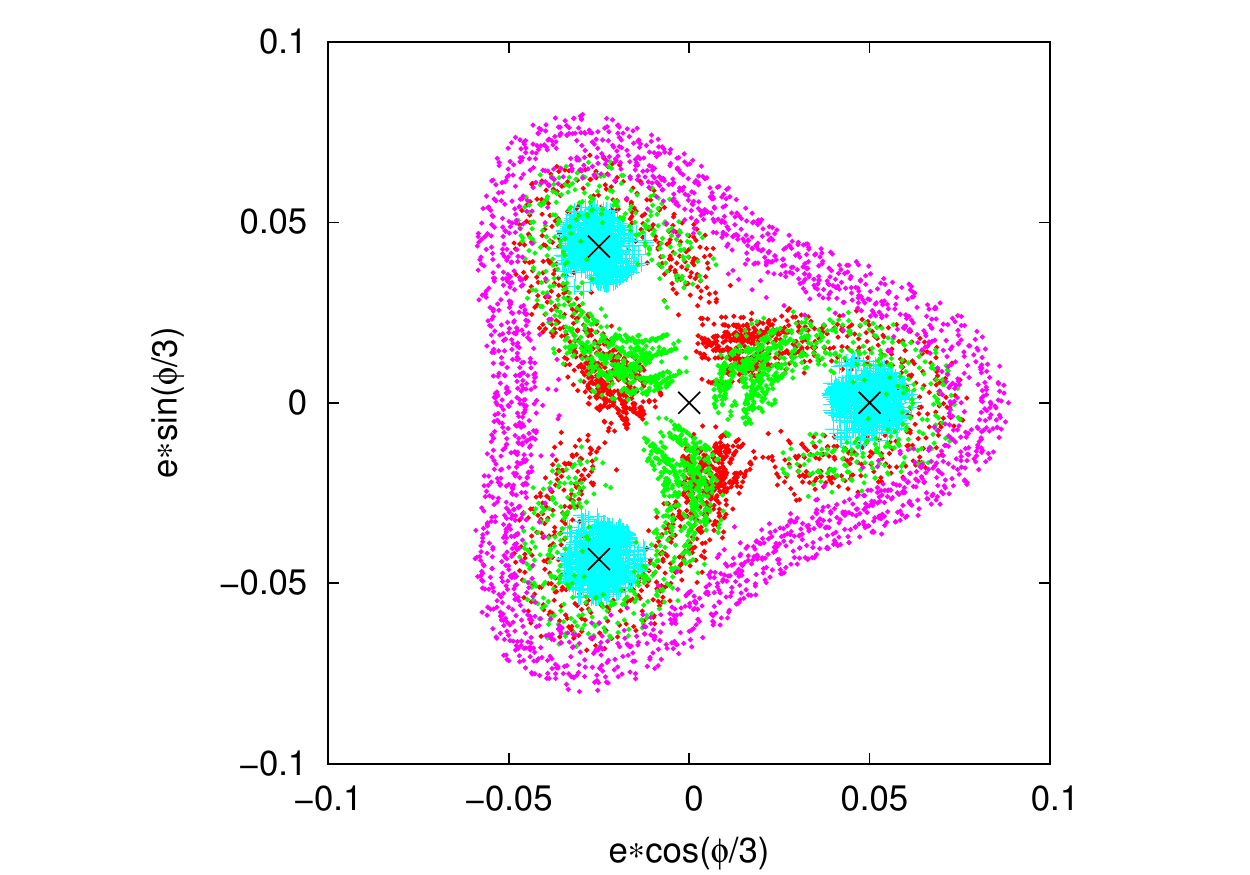}\label{hamilton:a}} \,
  \subfloat[]{\includegraphics*[width=6cm]{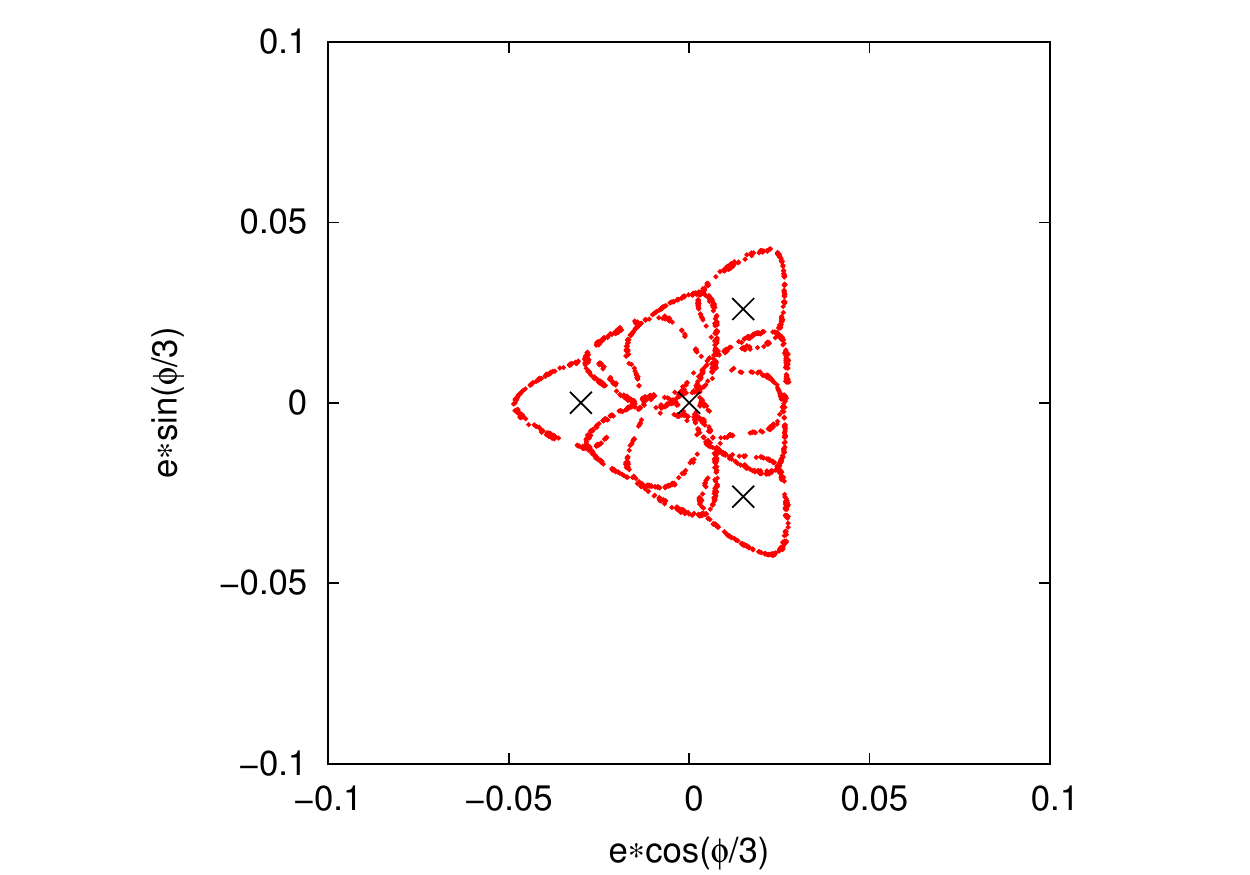}\label{hamilton:b}} \,
\caption{(a) Trajectories in the vicinity of 2/-1 resonance obtained by numerical integration of the equations of motion with  $\mu=0.01$ at $C=0.051$:  near exact resonance (cyan), separatrix (red and green) and outer circulation (magenta). (b) Trajectories in the vicinity of 1/-2 resonance obtained by numerical integration of the equations of motion with  $\mu=0.01$  at $C=-1.886$: near exact resonance in surface of section (red). The equilibrium points predicted by the analytic model are marked by crosses.}
\label{hamilton}
\end{center}
\end{figure*}

 
A resonant orbit corresponds to a set of 3 stable equilibrium points of the Hamiltonian (Eq.~\ref{hamiltonian}).
In Paper I we showed curves  of constant Hamiltonian  and location of equilibrium points  for several values of the parameter $\delta$. This is in agreement with the behavior observed in the surfaces of section near the 2/-1 and 1/-2 resonances (Fig.~\ref{sections}(b),(h)).  In particular, when $\delta=0$ (exact resonance) there is a bifurcation at the origin, and the 3 stable equilibrium points have $R=6$ and $\phi=0,\pm2\pi/3$  (2/-1 resonance) or $\phi=\pi,\pm\pi/3$  (1/-2 resonance). Applying the scaling we see that,  when $\mu=0.01$, the stable equilibrium points at the 2/-1 resonance have $e=0.05$, while the stable equilibrium points at the 1/-2 resonance have $e=0.03$.  In Fig.~\ref{hamilton} we show real trajectories in the vicinity of the 2/-1 and  1/-2 resonances obtained by numerical integration of the equations of motion with $\mu=0.01$ at Jacobi constant values ($C$) close to bifurcation at the origin.  There is very good agreement with the analytic model for the 2/-1 resonance (Fig.~\ref{hamilton:a}) since the equilibrium points and separatrix appear at the correct locations which implies that the scaling is correct. However, for the 1/-2 resonance an orbit close to the stable equilibrium points in the surface of section correspond to a large amplitude libration orbit in Fig.~\ref{hamilton:b}, possibly due to the vicinity of the separatrix at $e=0$.  It is also known that for exterior resonances it may be necessary to include higher order terms in the analytic model in order to obtain the correct dynamics \citep{Winter&Murray2}. What is important for our purposes is that the scaling is still approximately correct since the orbit encircles the equilibrium points predicted by the analytic model. This test provides further assurance on the correct identification of the resonant terms in Sect.~2.

\section{Discussion}
This article is the continuation of our work on retrograde resonances (Paper I). We identified the transformation that must be applied to the standard expansion of the three-dimensional disturbing function in order to obtain the relevant resonance terms for retrograde motion and analyze them quantitatively.

We  explored  the phase-space near the retrograde resonances 1/1, 2/1, and 1/2, by constructing surfaces of section for the planar circular restricted three-body problem using a mass ratio 0.01. The resonant term amplitude is of order $e^2$ for the 1/-1 resonance, while those of the  2/-1 and 1/-2 resonances are of order $e^3$. Therefore, these are the strongest retrograde resonances.

We saw that for low eccentricity non-crossing orbits,  libration in the 2/-1 resonance occurs around 0 (corresponds to starting at conjunction or opposition and pericenter) whereas libration in the 1/-2 resonance occurs around $180^\circ$ (corresponds to starting at conjunction and apocenter, or opposition and pericenter). These are the most stable configurations for non-crossing resonant orbits, since they ensure that closest approach with the secondary occurs always at pericenter for the 2/-1 resonance, and always at apocenter for the 1/-2 resonance.  
The behavior of low eccentricity 2/-1 and 1/-2 resonant orbits is in  reasonable agreement with an analytic model for 3rd order resonances based on the literal expansion of the disturbing function.
However, this analytic model is not valid for resonant crossing orbits. We observed that moderate to large eccentricity crossing orbits in the 2/-1 resonance or moderate eccentricity crossing orbits in the 1/-2 resonance may librate around 0 or $180^\circ$, while large eccentricity crossing orbits in the 1/-2 resonance may librate only around 0. 

Recalling that the literal expansion of the disturbing function is not adequate for co-orbital motion, we developed a semi-analytic model for the 1/1 retrograde resonance based on numerical averaging of the full disturbing function and valid for large eccentricity and inclination. We saw that this model correctly explains the 1/-1 resonant modes, namely libration around 0 and around $180^\circ$.  Whereas 1/-1 resonant libration around 0 is quite stable and occurs for a wide range of eccentricities, libration around $180^\circ$ occurs only near $e\approx 0$ and $e\approx 1$, hence it is located  close to the collision separatrix with the secondary. The disruptive effect of  these close encounters implies that 1/-1 libration around $180^\circ$ is easier to setup for smaller mass ratios ($\lesssim 10^{-4}$).  We also expect that transitions between different 1/1 retrograde resonant modes are possible in the three-dimensional problem, in analogy with what is described for prograde 1/1 resonant orbits \citep{Namouni1999,Namouni_etal1999}.

Shortly after completing this theoretical study of retrograde resonance we identified a set of Centaurs and Damocloids that are temporarily captured in retrograde resonance with Jupiter and Saturn \citep{Morais&Namouni2013}.
 
\section*{Acknowledgments}
We thank both reviewers for helpful suggestions that improved the article's clarity.
We acknowledge financial support from FCT-Portugal (PEst-C/CTM/LA0025/2011). The surfaces of section computations were performed on the Blafis cluster at the University of Aveiro.

\bibliographystyle{spbasic}   

\bibliography{CELE-D-13-00046}


\begin{figure*}[p]
\centering
\subfloat[]{\includegraphics*[width=9.5cm]{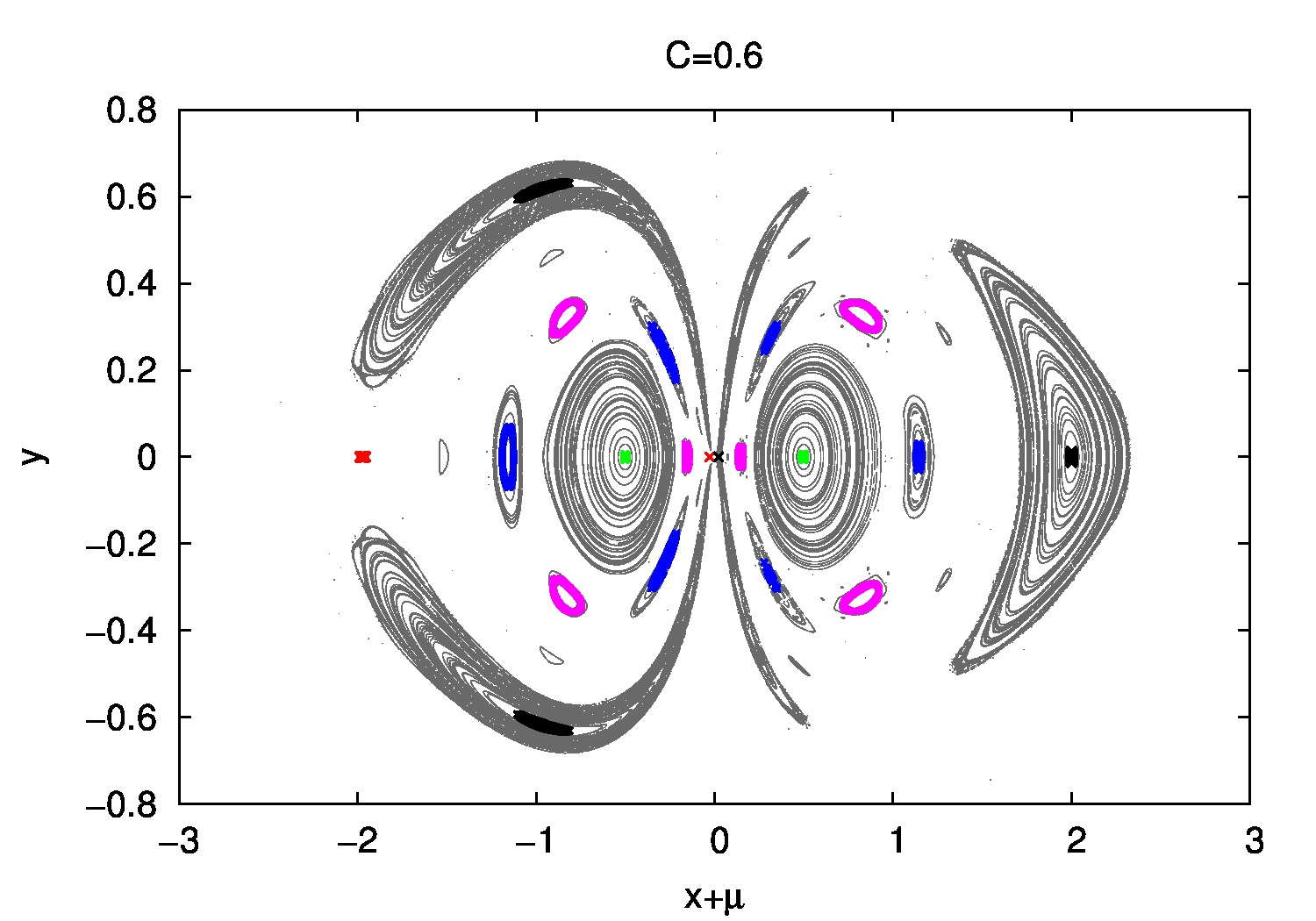}\label{sections:a} } \,
\subfloat[]{\includegraphics*[width=9.5cm]{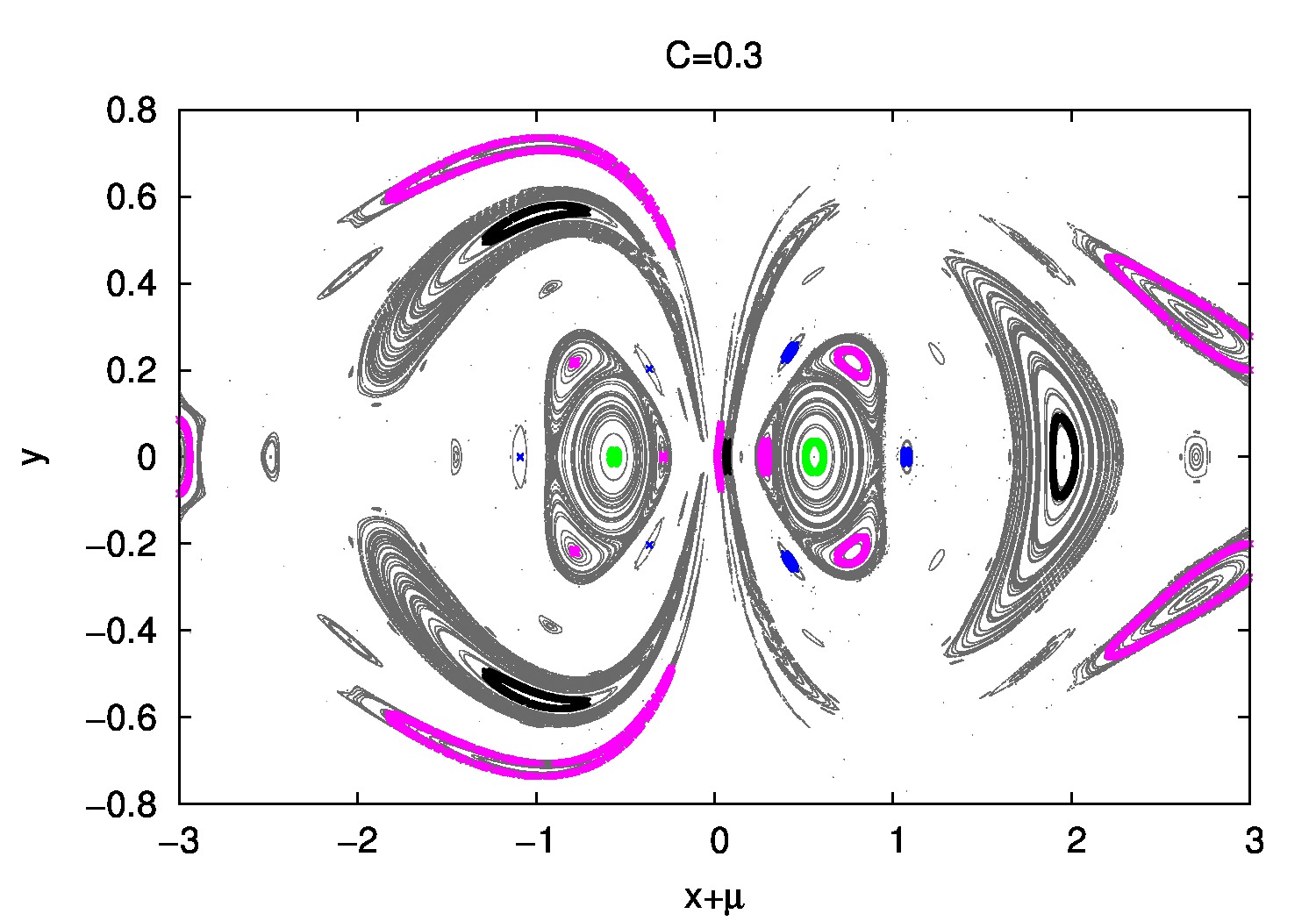}\label{sections:b} } \,
\subfloat[]{ \includegraphics*[width=9.5cm]{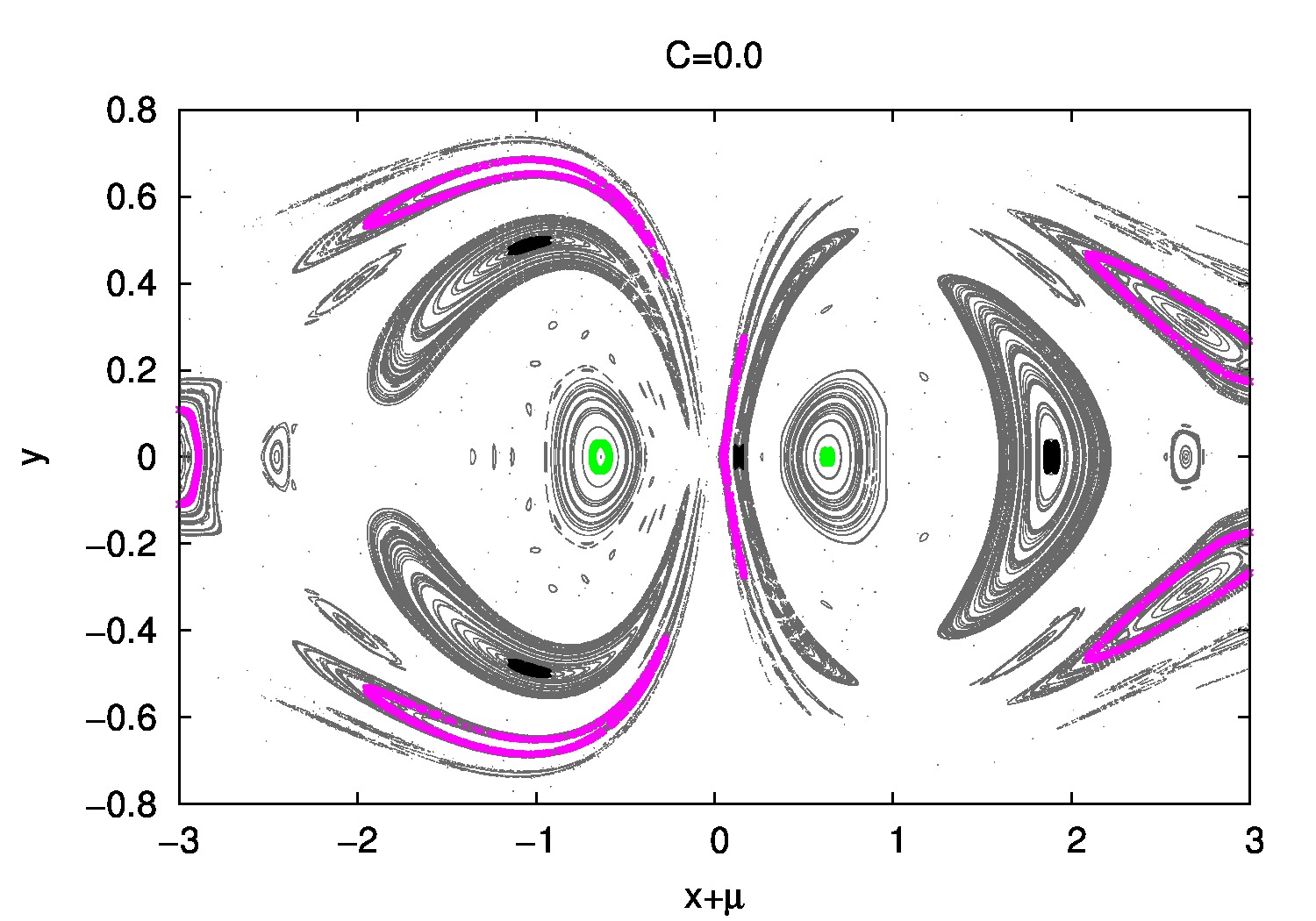}\label{sections:c} } \,
\caption[]{Surfaces of section for selected values of C. Primary is at $(0,0)$ and secondary is at $(1,0)$. Different colors correspond to different libration modes, or circulation (see text).}
\end{figure*}

\begin{figure*}[p]
\ContinuedFloat
\centering
\subfloat[]{\includegraphics*[width=9.5cm]{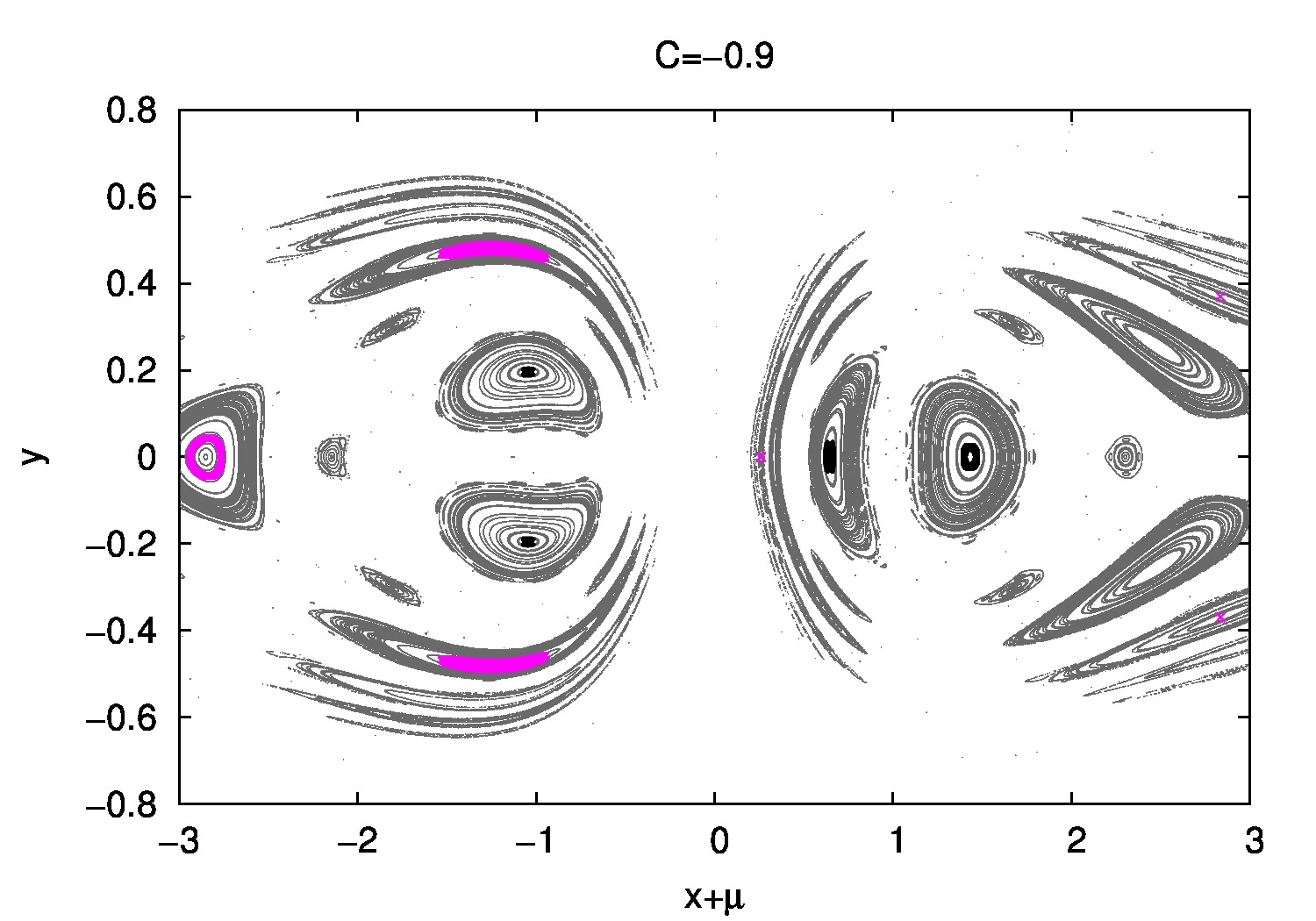}\label{sections:d} }  \,
\subfloat[]{\includegraphics*[width=9.5cm]{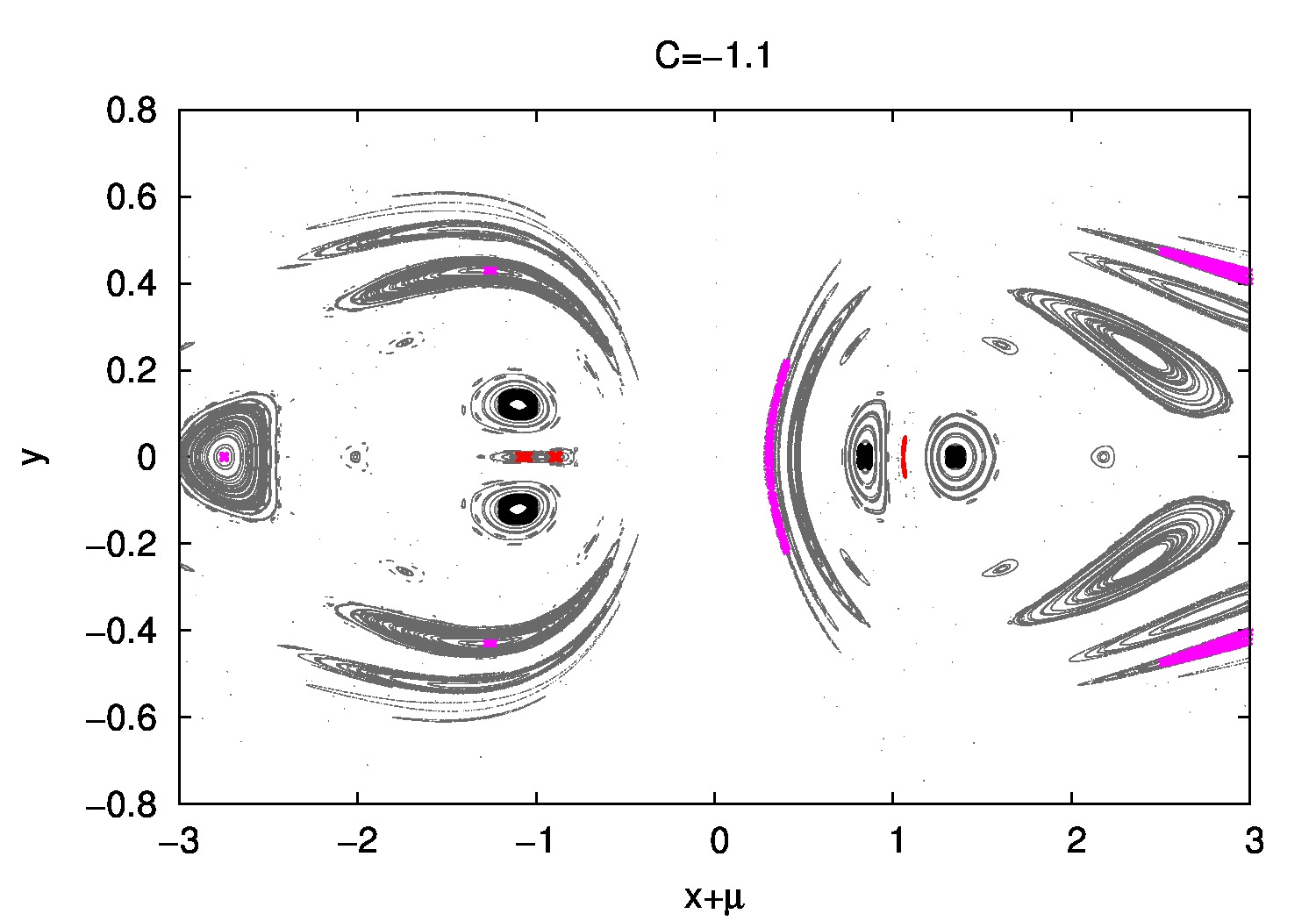}\label{sections:e} } \,
\subfloat[]{\includegraphics*[width=9.5cm]{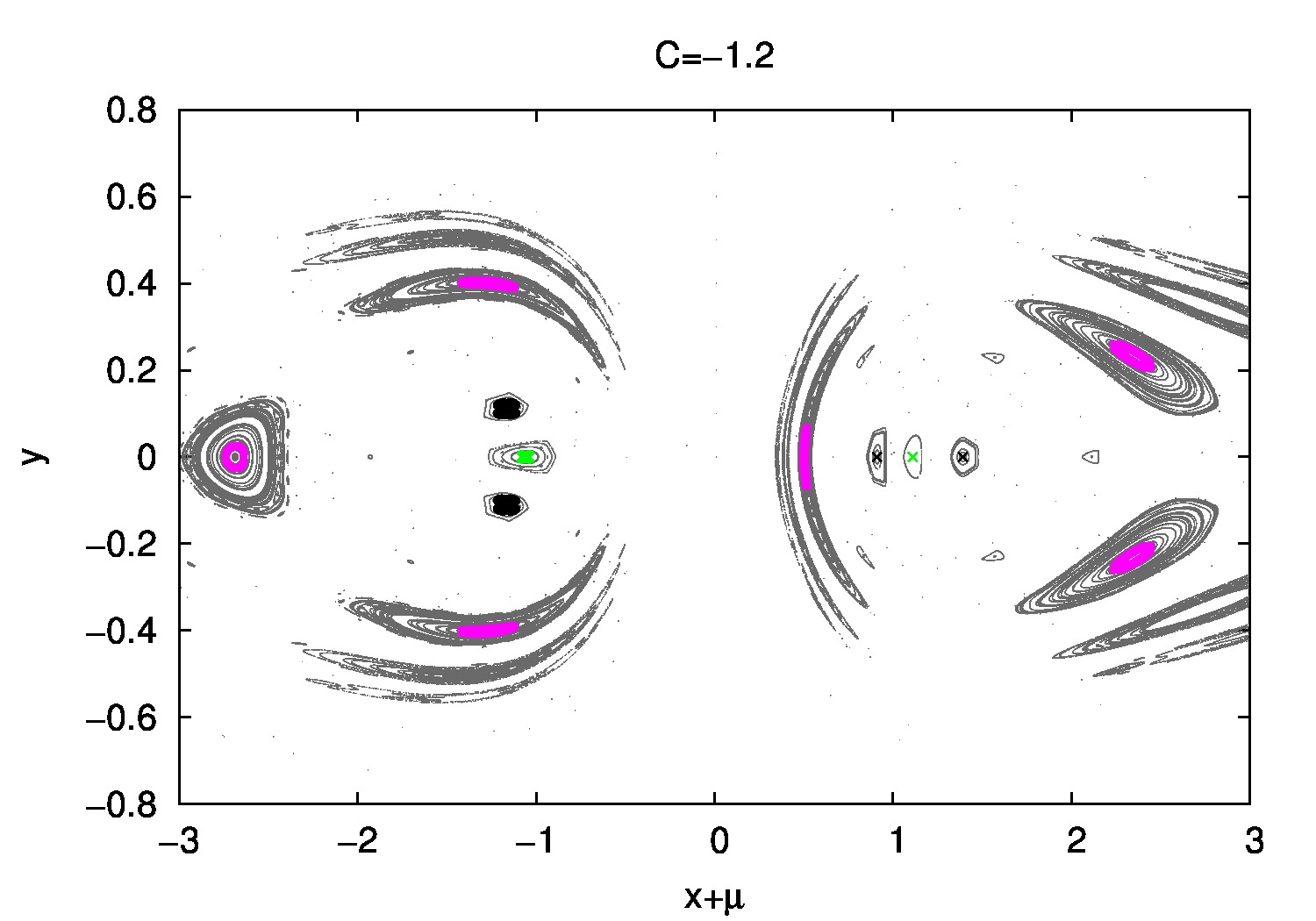}\label{sections:f} } \,
\caption[]{Surfaces of section for selected values of C. Primary is at $(0,0)$ and secondary is at $(1,0)$. Different colors correspond to different libration modes, or circulation (see text).}
\label{sections}
\end{figure*}

\begin{figure*}[p]
\ContinuedFloat
\centering
\subfloat[]{ \includegraphics*[width=9.5cm]{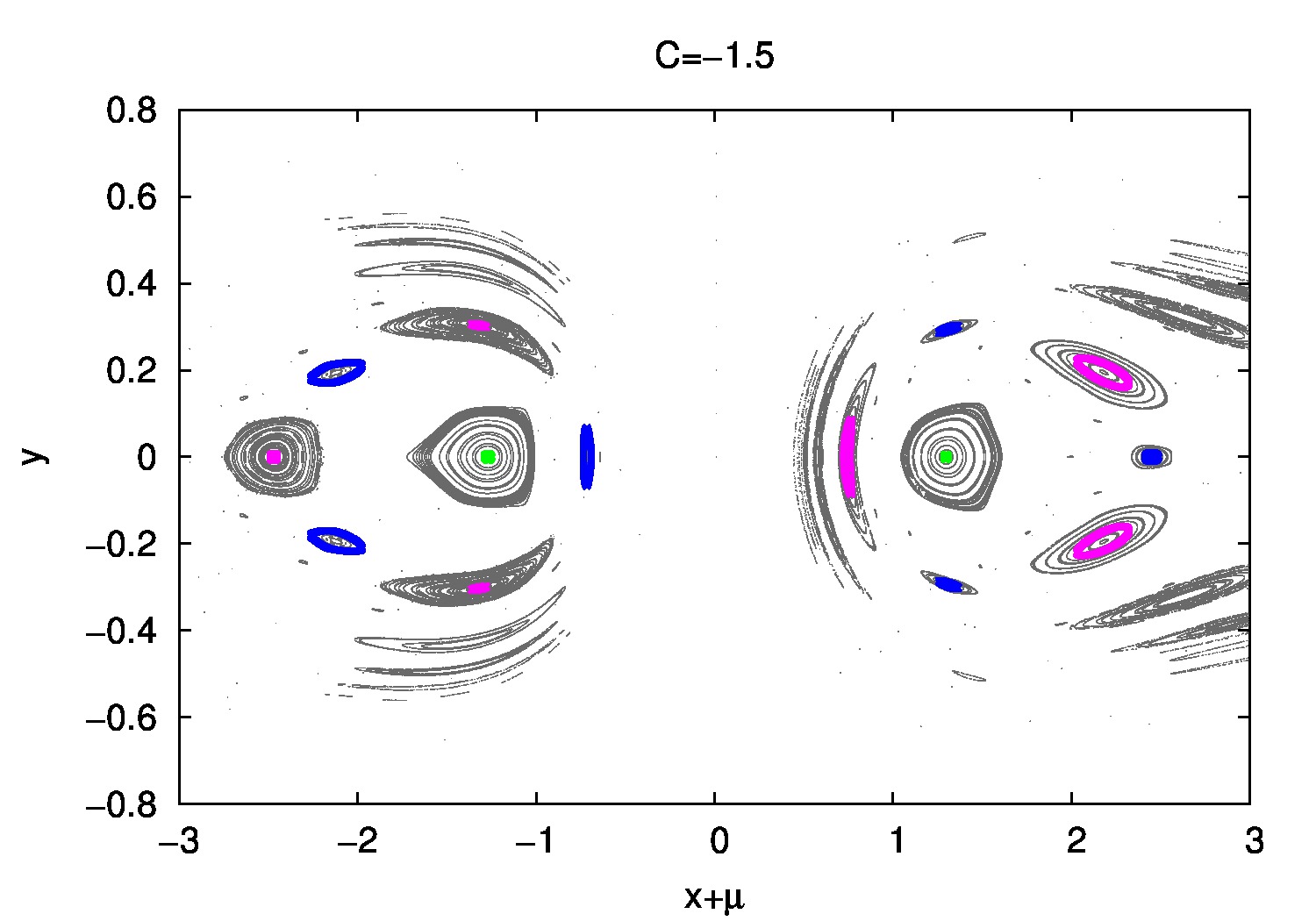}\label{sections:g} } \,
\subfloat[]{\includegraphics*[width=9.5cm]{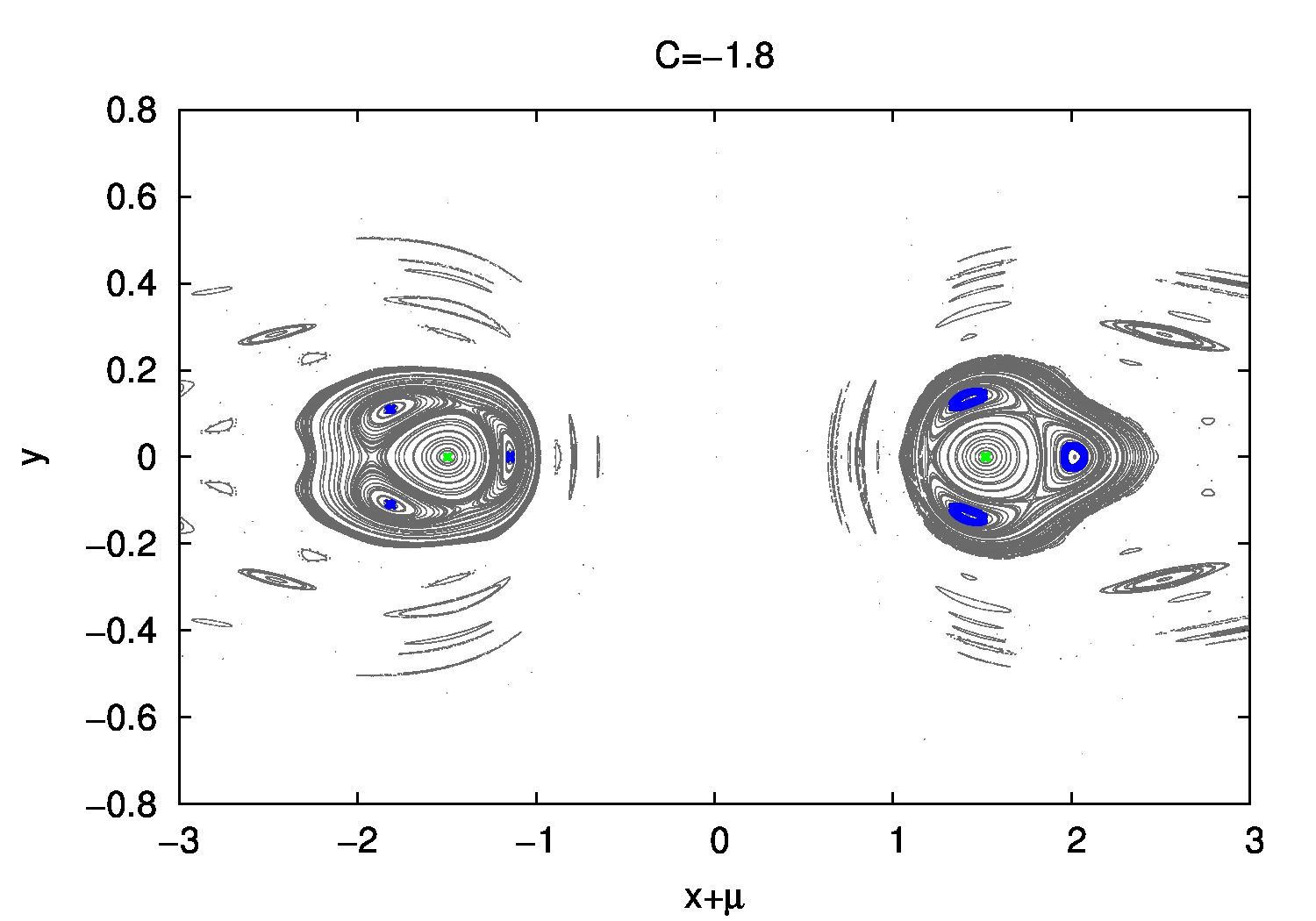}\label{sections:h} }  \,
\caption[]{Surfaces of section for selected values of C. Primary is at $(0,0)$ and secondary is at $(1,0)$. Different colors correspond to different libration modes, or circulation (see text).}
\label{sections}
\end{figure*}

\end{document}